\documentclass[superscriptaddress, twocolumn, pre, longbibliography, nofootinbib]{revtex4-1}
\usepackage{amsmath}
\usepackage{amsbsy}
\usepackage{amssymb}
\usepackage{graphicx}
\usepackage{color}
\usepackage{subfigure}
\usepackage{physics}
\usepackage{soul}
\usepackage{color}
\usepackage{bm}
\usepackage{array}
\usepackage{multirow}
\usepackage{lipsum}
\usepackage[normalem]{ulem}
\usepackage{verbatim}
\usepackage{natbib}
\usepackage{nccmath}
\usepackage[capitalize]{cleveref}
\begin{document}
\title{A strong non-equilibrium bound for sorting of crosslinkers on growing biopolymers}
%\title{A Thermodynamic Uncertainty Relation Bounds How Filament Polymerization Rate Modulates Molecular Sorting in Actin Bundling
%(A Strong Non-equilibrium Bound for Sorting of Biological Patterns on Growing Actin Bundling)}
\author{Yuqing Qiu}
\author{Michael Nguyen}
\affiliation{James Franck Institute, University of Chicago, Chicago, IL}
\affiliation{Department of Chemistry, University of Chicago, Chicago, IL}
\author{Glen M. Hocky}
\affiliation{Department of Chemistry, New York University, New York, NY}
\author{Aaron R. Dinner}
\author{Suriyanarayanan Vaikuntanathan$^*$}
\affiliation{James Franck Institute, University of Chicago, Chicago, IL}
\affiliation{Department of Chemistry, University of Chicago, Chicago, IL}
\begin{abstract}
Understanding the role of non-equilibrium driving in self-organization is crucial for developing a predictive description of biological systems, yet it is impeded by their complexity. The actin cytoskeleton serves as a paradigm for how equilibrium and non-equilibrium forces combine to give rise to self-organization. 
Motivated by recent experiments that show that actin filament growth rates can tune the morphology of a growing actin bundle crosslinked by two competing types of actin binding proteins [Freedman et al. PNAS 116, 16192–16197 (2019)], we construct a minimal model for such a system and show that the dynamics are subject to a set of thermodynamic constraints that relate the non-equilibrium driving, bundle morphology, and molecular fluxes.  The thermodynamic constraints reveal the importance of correlations between these molecular fluxes, and offer a route to estimating microscopic driving forces from  microscopy experiments. 
\end{abstract}
\maketitle

\section{Introduction}
Non-equilibrium driving is a crucial prerequisite for the function of many biological systems. Examples include kinetic proofreading~\cite{hopfield1974kinetic,Andrieux2008,Sartori2015,Poulton2019,murugan2012speed}, adaptation in molecular motors~\cite{seifert2011stochastic,murrell2015forcing,furthauer2019self}, and the suppression of phase decoherence in biochemical oscillators \cite{Lan2012,fei2018design,cao2015free,bryant2020energy,del2020high,del2020robust}, among others. Given the ubiquitous role played by non-equilibrium driving in biology, much recent work has been focused on establishing the general tradeoffs between energy consumption and organization~\cite{Joshi2017,Zhang2017,Tociu2019,Fodor2020,nguyen2016design,grandpre2020entropy}. Here, motivated by recent experimental work~\cite{Winkelman2016,freedman2019mechanical,bashirzadeh2020actin}, we consider growth and bundling dynamics of actin filaments and demonstrate that general energy-speed-morphology relations can be obtained for such systems. 

The actin cytoskeleton harnesses chemical energy to perform mechanical work that enables cells to migrate, divide, and exert forces on their surroundings, among other functions~\cite{mogilner1996cell,dickinson2009models,jegou68mechanically,murrell2015forcing,gardel2010mechanical,watanabe2008mdia2}. To perform these varied functions, a cell must be able to control the organization of its many components in both space and time. A growing body of evidence suggests that, surprisingly,  much of this organization can arise due to passive competition between actin binding proteins (ABPs) \cite{kadzik2020f}. 
At the same time, other processes such as the formation of a cytoskinetic ring require irreversible polymerization and motor activity \cite{vavylonis2008assembly,zimmermann2017mechanoregulated}. 
This suggests that cells can regulate their internal structures and, in turn, functions by tuning the relative contributions of passive and active processes.  Support for this idea comes from recent {\it{in vitro}} experiments and simulations that demonstrate that the the morphology of a growing actin bundle can be tuned not only by the binding affinities of the crosslinkers but also by the actin polymerization rates \cite{freedman2019mechanical}.

These observations, together with recent advances in non-equilibrium statistical mechanics~\cite{Tu2008,Tociu2019,murugan2012speed,barato2015thermodynamic,Gingrich2016,Lan2012}, raise the question whether the nonequilibrium driving---here due to polymerization---can be related to the emergent structure {\em quantitatively}.
Here, we address this question and present a theoretical framework that bounds the dynamics of a growing actin bundle. In particular, we derive constraints on a set of three matrices characterizing the process---a matrix containing the various non-equilibrium driving forces ($\boldsymbol \delta\boldsymbol \mu$, \cref{eq:centralMatdeltamu}), a matrix encoding the equilibrium and nonequilibrium morphologies ($\bf D$, \cref{eq:centralMatD}), and a matrix characterizing the covariance of the molecular fluxes ($\bf L^{-1}$, \cref{sec:constraints}). For these three matrices, we show that
\begin{equation}
    \label{eq:central1}
    \Tr{{\boldsymbol  \delta\boldsymbol \mu}-{\bf D}-{\bf L^{-1}}}\geq 0\,,{\rm Det}[{\boldsymbol  \delta\boldsymbol \mu}-{\bf D}-{\bf L^{-1}}]\geq0\,.
\end{equation} 
\cref{eq:central1} has a flavor of the fluctuation dissipation relation. Indeed, when the equality is satisfied, \cref{eq:central1} can be used to obtain a linear-response-like formula connecting the non-equilibrium forcing, the bundle morphology, and the response to fluctuations in molecular fluxes (\cref{eq:central22}). \cref{eq:central1} thus provides strong thermodynamic constraints on the non-equilibrium forcing, actin bundle morphology, and speed of growth. 
Notably, $\bf D$ and $\bf L^{-1}$ are experimentally accessible, such that \cref{eq:central1} can be used to bound $\boldsymbol \delta\boldsymbol \mu$, which is not straightforward to measure directly. 

%Further, it also provides a way to estimate the difficult to measure microscopic non-equilibrium driving forces, from measurements of the various fluxes and domain morphologies. 

 In what follows, we first outline a minimal model that captures the salient features of actin polymerization and bundling and show that it captures the observations desribed above. We then proceed to derive our central results, and show how these thermodynamic uncertainty relations constrain the dynamics.
 
\begin{figure}[tbp]
\centering
\includegraphics[scale=0.072,clip=true]{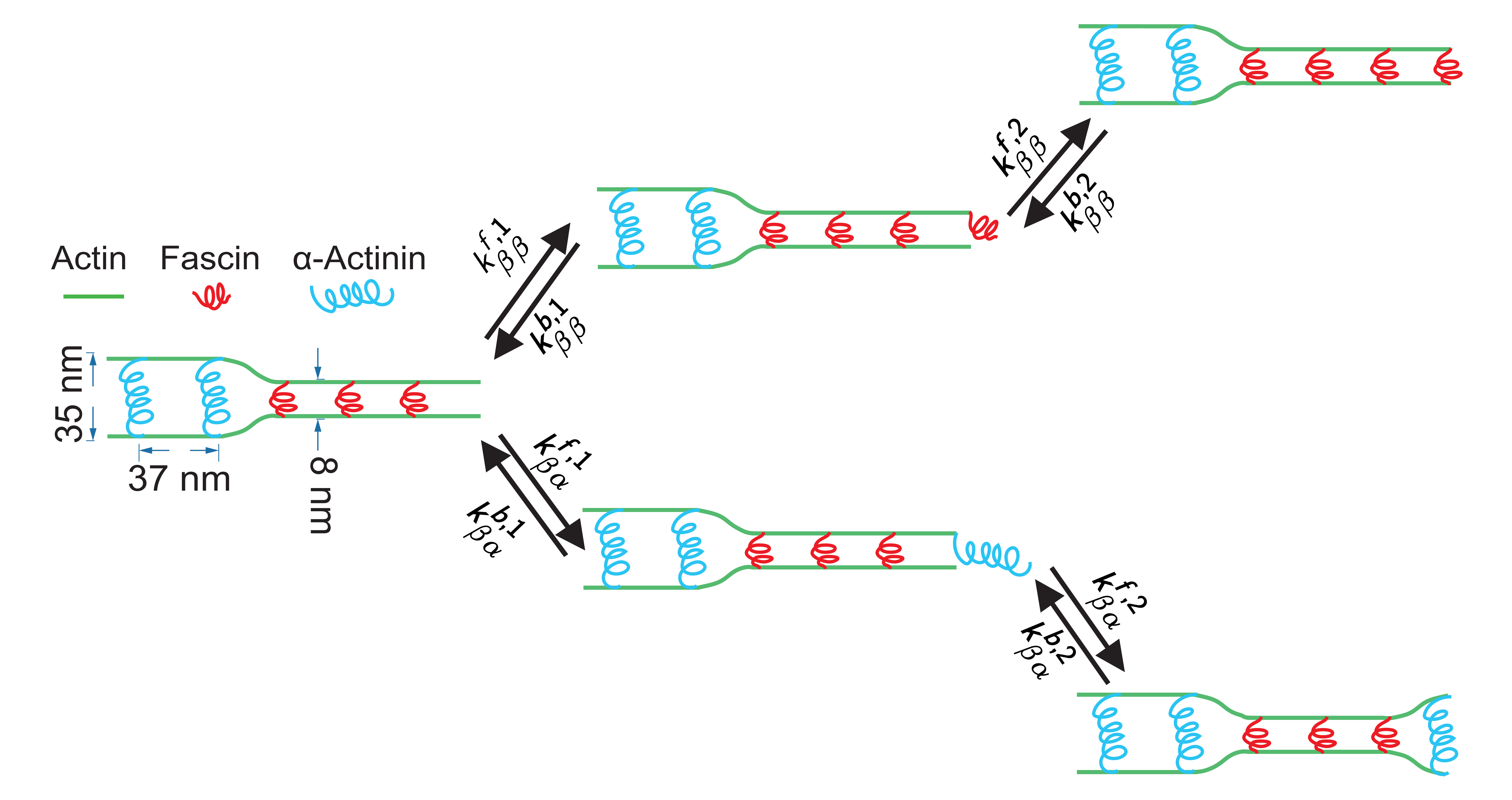}
\caption{ Schematic of adding one ABP at the tip of a growing actin bundle. Here, $\alpha$ and $\beta$ represent $\alpha$-actinin and fascin, which are 35 nm and 8 nm in size, respectively. The energetic cost of bending actin disfavors the binding a fascin after an $\alpha$-actinin, or {\it{vice versa}}, resulting in domains of consecutive $\alpha$ or $\beta$ types of ABPs. $k^{f,1}_{ij}$ and  $k^{b,1}_{ij}$ ($k^{f,2}_{ij}$ and  $k^{b,2}_{ij}$) are the forward and backward rates for the first (second) site of an ABP binding, where $i$ and $j$ are the types of the last two ABPs at the tip; $k^{f,1}_{\alpha\alpha} =k^{f,1}_{\beta\alpha}$  because both rates represent the binding of the first site of an $\alpha$ ABP. An analogous schematic can be drawn for the case that the second-to-last ABP is $\alpha$, and this case introduces four additional pairs of forward and backward rates with corresponding constraints.}
\label{fig:bundlescheme}
\end{figure}

\section{Actin polymerization drives sorting of actin bundling proteins}
Inspired by the experiments in Refs.~\onlinecite{Winkelman2016,freedman2019mechanical}, we consider a bundle consisting of two parallel actin filaments and two types of ABPs, $\alpha$ and $\beta$.  The growth of a parallel actin bundle involves continuous actin monomer addition at one end, as well as continual binding and `zipping' of the bundle by ABP binding at that same end. 
In the specific case shown in \cref{fig:bundlescheme}, $\alpha$ and $\beta$ represent crosslinking proteins $\alpha$-actinin and fascin, respectively, such that the bundles formed by the $\alpha$ ABPs are substantially more widely spaced than those formed by the $\beta$ ABPs. Consequently, the bending penalty of actin implicitly favors addition of the current crosslinker at the growing end, as it costs energy to switch from one type to the other. 
Under conditions of equilibrium, the cost of bending actin favors the formation of distinct domains of only $\alpha$ or $\beta$ ABPs. 
In the case of {\em in vitro} experiments using an equimolar mixture of fluorescently labeled $\alpha$-actinin and fascin binding to growing actin bundles, these domains are on the order of several micrometers ($\sim$ 100 crosslinkers) long \cite{Winkelman2016}.
As noted above, it has been reported~\cite{freedman2019mechanical} that the domain length statistics can be modulated by the rate of actin polymerization. 
%Below, we show that the statistics of domain length, and hence the composition and morphology of the actin bundle, can in fact be constrained using results derived from non-equilibrium thermodynamics. 

We construct a minimal model of this system with the following simplifications: first, we assume that the actin binding proteins can only bind and unbind from the sites at the leading edge and not from the bulk of the actin filament, and second, we assume that the two binding sites of each ABP bind sequentially and do not allow an ABP to bind to a single filament with both its sites.  As a result, we need two pairs of forward rates $k^{f,1}_{ij}$ and backward rates $k^{b,1}_{ij}$ to describe the binding of first site of each ABP and two other pairs, $k^{f,2}_{ij}$ and $k^{b,2}_{ij}$, for their second site (\cref{fig:bundlescheme}).  The consideration of both heads independently is more sophisticated than the kinetic Monte Carlo (KMC) models considered in Refs.~\onlinecite{Winkelman2016,freedman2019mechanical} and is consistent with the experimental and simulation observations therein.

Here, we further decompose the forward rate of $k^{f,m}_{ij}$  into an equilibrium component, $k^{f,m}_{ij,eq}$, that satisfies a local detailed balance rule and accounts for all the energetics associated with ABP binding and filament deformations, and a component $dk^{f,m}_{ij}$ that can model any non-equilibrium contributions to the rate,
\begin{equation} \label{eq:ratesk}
\begin{split}
k^{f,m}_{ij} = k^{f,m}_{ij,eq} dk^{f,m}_{ij}. 
\end{split}
\end{equation}
%We assert that only the forward rates are affected by actin polymerization, since this is a consequence of extending the flexible area available for new crosslinker binding \cite{freedman2019mechanical}, and hence set all the $k^{b,m}_{ij}$ to unity. 
We assert that only the forward rates are modified by any non-equilibrium effects including actin polymerization and we set all the $k^{b,m}_{ij}$ to unity.  
The equilibrium factor $k^{f,m}_{ij,eq}$ accounts for the binding affinity of an ABP. In cases where an attached ABP binds to the second actin filament, this equilibrium part also accounts for the energy penalty associated with bending the actin filament if the newly bound ABP is different from the previous ABP at the tip ({e.g.,} rate $k^{f,2}_{\beta\alpha,eq}$ in \cref{fig:bundlescheme}) and the free energy associated with zipping the actin bundle ({e.g.,} rates $k^{f,2}_{\beta\beta,eq}$ and $k^{f,2}_{\beta\alpha,eq}$ in \cref{fig:bundlescheme}). 

The non-equilibrium component in our model, $dk^{f,m}_{ij}$, heuristically accounts for any effects due to the finite rate of actin growth and polymerization, excess concentration or chemical potential of various ABPs in solution, and their molecular structure. Given this, we generically decompose the non-equilibrium components as 
\begin{equation} \label{eq:driving-in-rates-dk}
\begin{split}
dk^{f,m}_{ij} = dk_{j} =1+ f_{{\rm molecular},j} f_{{\rm density},j} f_{{\rm pol},j}  \\
%dk^{f,m}_{i\alpha} = dk_{\alpha} =1+ f_{\rm molecular,\alpha} f_{\rm density,\alpha} f_{\rm pol,\alpha}  \\
%dk^{f,m}_{i\beta} = dk_{\beta} = 1+ f_{\rm molecular,\beta} f_{\rm density,\beta} f_{\rm pol,\beta} 
\end{split}
\end{equation}
where $i$ and $j$ are the types of ABPs at the bundle tip. The factor $f_{\rm pol}$ models the modulation of the rates due to the finite rate of growth of the actin filaments ($k_\mathrm{grow} )$. Specifically, over a time scale $\tau$, the average increase in the number of binding sites on the filaments is $k_\mathrm{grow}\tau$ and \(k^{f,1}_{ij,eq}\tau\) is the number of binding events per binding site. Assuming Poisson statistics, the net rate of ABP binding is hence modulated by the factor, 
\begin{equation}\label{eq:polymerization-driving-fpol}
\begin{split}
f_{{\rm pol},i} = 1 - e^{- k^{f,1}_{ii,eq} k_{\rm grow}\tau^{2}}
%f_{\rm pol,\alpha} = 1 - e^{- k^{f,1}_{\alpha\alpha,eq} k_{\rm grow}\tau^{2}} \\
%f_{\rm pol,\beta} = 1 - e^{- k^{f,1}_{\beta\beta,eq}  k_{\rm grow}\tau^{2}}
\end{split}
\end{equation}
The factor $f_{\rm pol}$ is essentially the probability of binding at least one ABP within time $\tau$. $f_{\rm pol}$ increases from a value of $0$ when $k_{\rm grow}$ is negligible to a value of $1$ for rapid actin polymerization. It acts as a scaling factor that tunes the rates from their equilibrium values $k^{f,m}_{ij,eq}$ to their maximum rates. 
%$(1+f_{\rm molecular,j} f_{\rm density,j})k^{f,m}_{ij,eq}$

The rates of ABP binding are also influenced by the ABP concentrations in solution around the actin filaments. The phenomenological factor $f_{\rm density}$ accounts for these effects. Finally, the phenomenological factor, $f_{\rm molecular}$ has been introduced to account for any remaining kinetic differences between the ABPs. Such factors could modulate the maximum rates of adding ABPs in fast growing bundles, but they do not affect the equilibrium rates $k^{f,m}_{ij,eq}$ and the corresponding equilibrium structure of the bundle \cite{freedman2019mechanical}. 

%For instance, an ABP that is long in size and has floppy heads, such as $\alpha$-actinin, is expected to have a longer binding timescale in order to position itself between the parallel actin filaments, in relation to the binding timescale of a compact ABP, such as fascin.nd assigning a relative lower molecular factor for $\alpha$-actinin than for fascin

\begin{figure}[tbp]
\includegraphics[scale=1,trim= 0cm 0cm 0cm 0cm, clip=true]{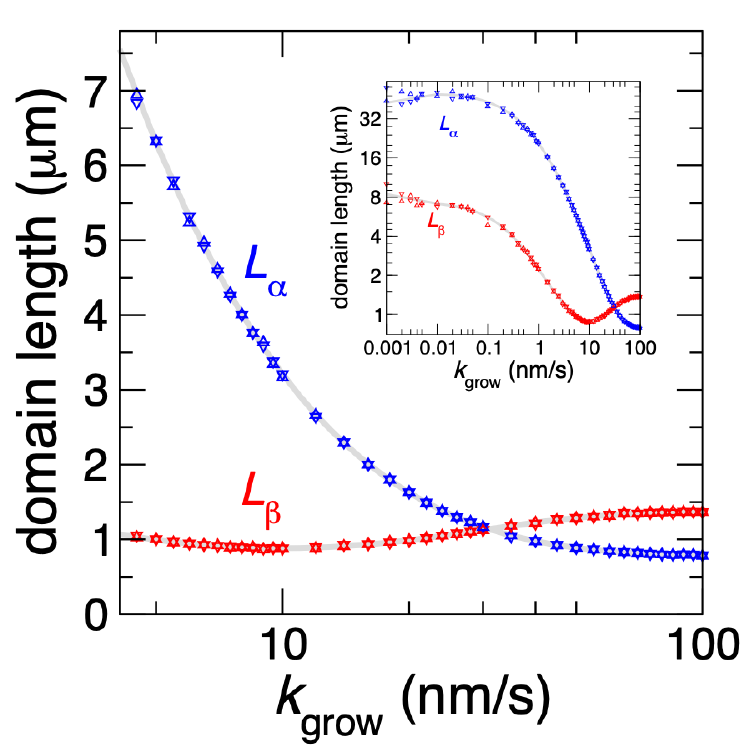}
\caption{Domain length of $\alpha$ and $\beta$ ABPs as a function of polymerization rate $k_\mathrm{grow}$. The distance between neighboring ABPs is assumed to be 0.037 $\mu$m in computing domain lengths \cite{Winkelman2016}. Blue and red triangles are domain lengths $L_{\alpha}$ and $L_{\beta}$ measured from KMC simulations. Up and down triangles represent the domain lengths measured in simulations with initial configurations composed of either all $\alpha$ or all $\beta$ types of ABPs, respectively. Gray lines (\cref{eqs14}) are the domain lengths computed by self-consistently solving the master equation (\cref{eq:master-eq}). The parameters for both KMC simulations and the master equation are $k^{f,1}_{\alpha\alpha,eq} = 6$, $k^{f,1}_{\beta\beta,eq} = 2$, $f_{\rm density,{\alpha}}=f_{\rm density,{\beta}} = 100$, $f_{\rm molecular,\alpha} = 0.4$, $f_{\rm molecular,\beta} = 1$, $L_{\alpha,\rm eq} = 900 (33.3 \mu$m), $L_{\beta, \rm eq} = 300 (11.1 \mu$m) and $\tau =\rm 1 s$. The inset shows the domain lengths over a wider range of polymerization rates with the same symbols. The  plateaus toward the left of the inset represent the domain lengths approaching their equilibrium values.}
\label{fig:domain-lengths}
\end{figure}

KMC simulations of this minimal model (described in SI Sec.~\ref{SI:KMC}) reproduce the crossover of domain lengths in fast growing actin bundles (\cref{fig:domain-lengths}) \cite{freedman2019mechanical}. 
%Equilibrium detailed balance dynamics are achieved when the $dk^{f,m}_{i\alpha/\beta}$ rates are set to unity. 
%Our model shows how the molecular properties of ABPs can determine their abilities to secure binding sites on the growing actin filaments and thus harness the non-equilibrium driving forces provided by actin polymerization (\textbf{SI Fig. 4}).  
It thus captures the essential physics of the system 
%While our miminal model provides insight into how the kinetics of actin polymerization can tune the ABP domain sizes, our treatment so far does not show how these non-equilibrium effects are constrained thermodynamically.  
and serves as a meaningful starting point for development of a theoretical framework that shows that the behavior is bounded by a general energy-speed-morphology relation.

\section{Connections between the growth and morphology of actin bundles: A Markov state model}

\begin{figure}
\centering
\includegraphics[scale=0.3,trim=18cm 10cm 8cm 10cm, clip=true]{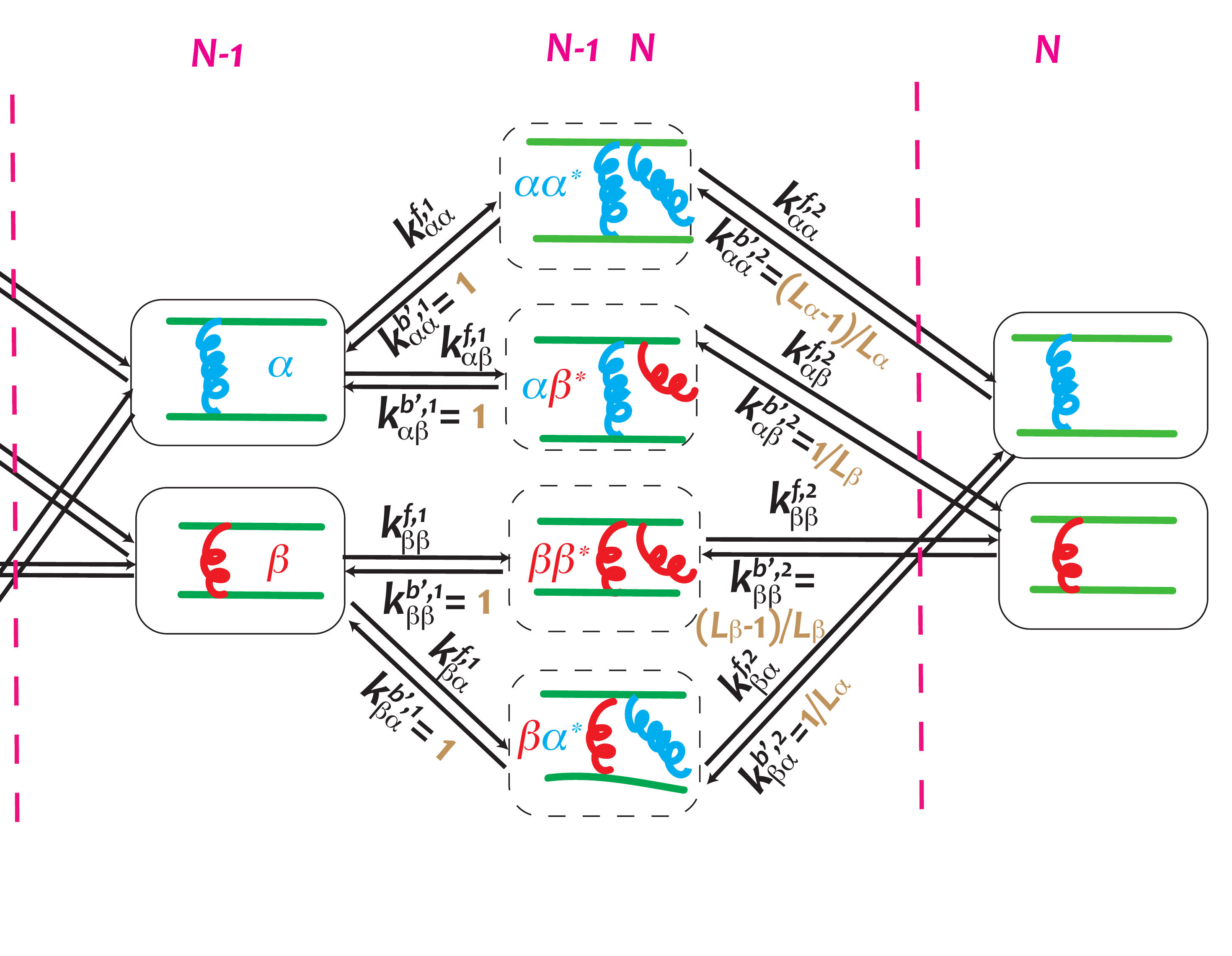}
\caption{Schematic of the $N^{th}$ cross-linker binds to the actin pairs. The states between two pink dashed lines describe the $N^{th}$ ABP binding. Arrows link states between which the transitions are allowed. $k^{f,m}_{ij}$ and $k^{b,m}_{ij}$ are the forward and backward rates, where $i$ and $j$ are the types of the last two ABPs at the tip and $*$ represents a half bound state. The rates used in the KMC simulations and in the master equation are compared in \cref{tables1}.}
\label{fig:kmc-scheme}
\end{figure}

The number of accessible states of the Markov chain for actin polymerization and bundling, as defined in the last section, grows rapidly as a function of time. Writing down thermodynamic relations for such growing systems becomes cumbersome. Here, we show that it is in fact possible to account for the behavior of the growing system using only a tractable, finite-state, Markov model. Using this model, we derive thermodynamic bounds for the non-equilibrium sorting process in \cref{fig:kmc-scheme}. 

We begin by introducing a mean-field treatment for the various configurations that arise at the tip as ABPs associate and dissociate sequentially (\cref{fig:kmc-scheme}). The forward rates $k^{f,m}_{ij}$ are consistent with those used in KMC simulations, accounting for the energetic terms in binding ABPs and the effect of actin polymerization.  
%composed of the equilibrium part $k^{f,m}_{ij,eq}$ derived from the energetic terms and the driving part derived from the actin polymerization rate and properties of ABPs, The forward R rates depend on [explain]. The forward and backward w rates are [explain/refer back to the previous section where you would have explained them!]. 
%[Talk about how they are derived from energetics etc if you haven't already] . [Talk about how everything is consistent with KMC]. 
The backward rates in the finite-state model self-consistently account for the probability of finding the appropriate ABP in the bulk of the actin bundle. For instance, using $L_{\alpha}$ to denote the domain lengths of $\alpha$ type ABPs and $L_{\beta}$ for its $\beta$ counterpart, the probability of finding an $\alpha\alpha*$ configuration (where $*$ stands for half bound state as in \cref{fig:kmc-scheme}) during unbinding one head of the last $\alpha$ ABP is $(L_{\alpha}-1)/L_{\alpha}$ and the probability of finding a $\beta\alpha*$ configuration is $1/L_{\alpha}$. In computing the unbinding rate $k^{b',m}_{ij}$ used in the finite-state model, we multiply the backward rates $k^{b,m}_{ij}$ in KMC simulations by the corresponding conditional probabilities. %modify this following sentence
Note that the effective backward rate for unbinding the first head of an ABP, $k^{b',1}_{ij}$, is equal to $k^{b,1}_{ij}$ in KMC simulations because the conditional probabilities in these transitions are 1. The difference in  $k^{b',1}_{ij}$ and $k^{b',2}_{ij}$ is because the unbinding of an ABP's first site $k^{b',1}_{ij}$ initiates from state $ij*$ and reaches state $i$, so the type of the preceding ABP is known from the current state, while the unbinding of ABP's second site $k^{b',2}_{ij}$ initiates from state $j$ and the type of the preceding ABP $i$ is uncertain (see \cref{fig:kmc-scheme}). We confirm the expressions of these conditional probabilities by extracting the effective backward rates and domain length ratios from the full KMC simulations (\cref{figs2}). 
%To test whether the effective backward rates $k^{b',m}_{ij}$ restore the lost information due to the finite number of states considered in master equation, we count the statistics of finding the appropriate ABP in KMC simulations and confirm that they are indeed proportional to the domain lengths measured from the corresponding bundle structure (\textbf{Fig 2 in SI}). The high quality of this approximation in computing $k^{b',m}_{ij}$ is not surprising because the parameter range we explore in this study always results in domain lengths much longer than 1, {\it{i.e}}. $L,S >> 1$. This ensures the insensitivity of the conditional probability in unbinding to the fluctuations in domain lengths. %Second, the final configurations contains a large number of alternations between the two domains, so that the average domain lengths converge to steady states. 

Given these expressions for the rates, we can write a master equation describing the evolution of probabilities of the various {\it tip} configurations, $P_{ij}$:
\begin{equation} \label{eq:master-eq}
\begin{split}
WP =\sum_{i,j=\alpha,\beta}(&k^{f,1}_{ij}P_{i} - k^{b',1}_{ij}P_{ij*}\\ +\ &k^{f,2}_{ij}P_{ij*} - k^{b',2}_{ij}P_{j} ) = 0.
\end{split}
\end{equation}
However, this master equation depends on the average domain length $L_{\alpha}$ and $L_{\beta}$. To solve the master equation of this system and get closed form expressions of $P_{ij}$ at the steady state, we require a relation that connects the tip configuration probabilities to the domain lengths. Such a connection can be obtained by noting that as the bundle barbed end grows, the tip configuration merges into the bulk of the bundle. The probabilities of tip configurations at steady state together with their corresponding rates determine the relative amounts of the two types of ABPs growing into the bulk. In other words, the fluxes $ J_{N-1,i,N,j} = k^{f,1}_{ij}P_{i} - k^{b',1}_{ij}P_{ij*} = k^{f,2}_{ij}P_{ij*} - k^{b',2}_{ij}P_{j}$ are proportional to the probabilities of sampling the corresponding ABPs in the bulk (see \cref{eqs10}). 

%contain both of these information ), so we should be able to constrain the domain length ratio $L/S$ in the bulk using these currents. 

Indeed, this reasoning can be put on a firm mathematical footing by adapting the calculations in Ref.~\onlinecite{gaspard2014kinetics}. Specifically, as described in SI Section~\ref{SI:master}, we derive self-consistency conditions that relate the currents at which various tip configurations grow to the domain lengths in bulk: 
 \begin{equation}
\begin{split}\label{eq:mean-field-currents-domains}
J_{N-1,\alpha,N,\alpha}&= \frac{L_{\alpha}-1}{L_{\rm tot}}J_{\rm tot}\\
J_{N-1,\beta,N,\beta}&= \frac{L_{\beta}-1}{L_{\rm tot}}J_{\rm tot}\\
J_{N-1,\alpha,N,\beta}=J_{N-1,\beta,N,\alpha} &= \frac{1}{L_{\rm tot}}J_{\rm tot}\\
\end{split}
\end{equation}
where $N-1,i,N,j$ denotes the tip configurations at position $N-1$ and $N$, $ J_{\rm tot}$ is the sum of the four currents, and $L_{\rm tot} \equiv L_{\alpha} + L_{\beta}$. We measure the currents and domain lengths from KMC simulations, and \cref{figs3} demonstrates the validity of the relations between the fluxes and domain lengths in \cref{eq:mean-field-currents-domains}.
%Explain the currents somewhere around here .. either before or after the equation .. you can simplify notation by using somethign like  $J^{\delta N=1}_{ij}$ etc if possible ... make it as easy to read and digest as possible .. the physical picture should come out very clearly ! ]
%[NEED to REWRITE THIS .. NEEDS TO BE MUCH BETTER EXPLAINED ! ]

The master equation (\cref{eq:master-eq}) can now be solved with these additional self-consistency conditions. Expressions for the non-equilibrium domain lengths \(L_{\alpha}\) and \(L_{\beta}\) can also be readily obtained (\cref{eqs14}). The gray lines in \cref{fig:domain-lengths} illustrate that these predictions are in excellent agreement with the domain lengths in KMC simulations for the full range of actin polymerization rates. This model also recovers the trend shown previously by simulations \cite{freedman2019mechanical} that as the binding affinity of the short crosslinker (equivalent to $\beta$ ABP in our model) is weakened, the crossover of domain lengths is deferred to a faster growth speed. Thus our mean-field treatment is able to capture the behavior of the model quantitatively. 

%[MOVE FIG 3 +related discussion TO SI, WE CAN JUST WORK WITH FIG 1 in the MAIN TEXT, WE CAN QUICKLY SAY HOW VARIOUS TRENDS ARE RECOVERED].
%[Talk about how the KMC and master equation are working well .. ...you can put some graphs here from Section III .. Section III can just be a discussion]
%To probe the thermodynamic bound for a growing actin bundle, we first write down the rate of entropy production and compute this quantity for the Markov chain. In section 3.3, we compare this exact entropy production rate with a thermodynamic bound derived from the kinetic information in simulations. This enables us to test the idea of constraining the allowed patterns in a non-equilibrium growth process with minimal kinetic information. 
\section{Thermodynamic constraints between the non-equilibrium forcing, fluctuations, and morphology}
\label{sec:constraints}
The non-equilibrium thermodynamics of the growing actin bundle can now be probed. Using the master equation (\cref{eq:master-eq}) and the finite-state Markov model in \cref{fig:kmc-scheme}, the entropy production rate $\dot\sigma$ for our effective Markov model can be written as
\begin{equation} 
\begin{split}\label{eq:EP-secondlaw}
\dot\sigma = J_{\rm tot} (\Delta\mu   - \varepsilon_{diss})\geqslant 0.
\end{split}
\end{equation}
The factor $\Delta \mu$ represents the non-equilibrium forces driving polymerization; it is defined as 
\begin{equation}
    \label{eq:deltamu}
    \Delta\mu = \frac{2}{L_{\rm tot}}\left(\sum_{i}^{\alpha,\beta} {L_{i}\log{dk_{i}}} \right).
\end{equation}
The factor $\varepsilon_{diss}$ is a measure of the difference between the non-equilibrium and equilibrium morphologies as charecterized by the respective average domain lengths $L_{\beta, \rm eq}$ and $L_{\alpha, \rm eq}$. 
\begin{equation}\label{eq:ediss}
\begin{split}
\varepsilon_{diss}=-\frac{1}{L_{\rm tot}}&\bigg (\sum_{i}^{\alpha,\beta} {L_{i}}
\log\frac{L_{i}}{L_{i, \rm eq}}
\\
&-
\sum_{i}^{\alpha,\beta} ({L_{i}-1})
\log\frac{L_{i}-1}{L_{i, \rm eq}-1} \bigg)\,.
\end{split}
\end{equation}
\cref{eq:EP-secondlaw} is a statement of the second law of thermodynamics.  However, we can improve on this bound substantially by adapting recent work \cite{yan2019achievability,dechant2018multidimensional}. Specifically, we show in the SI Section.~\ref{SI:FDT} that a stronger matrix relation can be obtained that is valid far from equilibrium. This is our main result, \cref{eq:central1}, which we reproduce here for convenience:
\begin{equation}
    \label{eq:central11}
   \Tr{{\boldsymbol \delta\boldsymbol\mu}-{\bf D}-{\bf L^{-1}}}\geq 0\,,{\rm Det}[{\boldsymbol \delta\boldsymbol\mu}-{\bf D}-{\bf L^{-1}}]\geq0\,.
\end{equation}

\begin{figure}
\includegraphics[scale=0.40, clip=true]{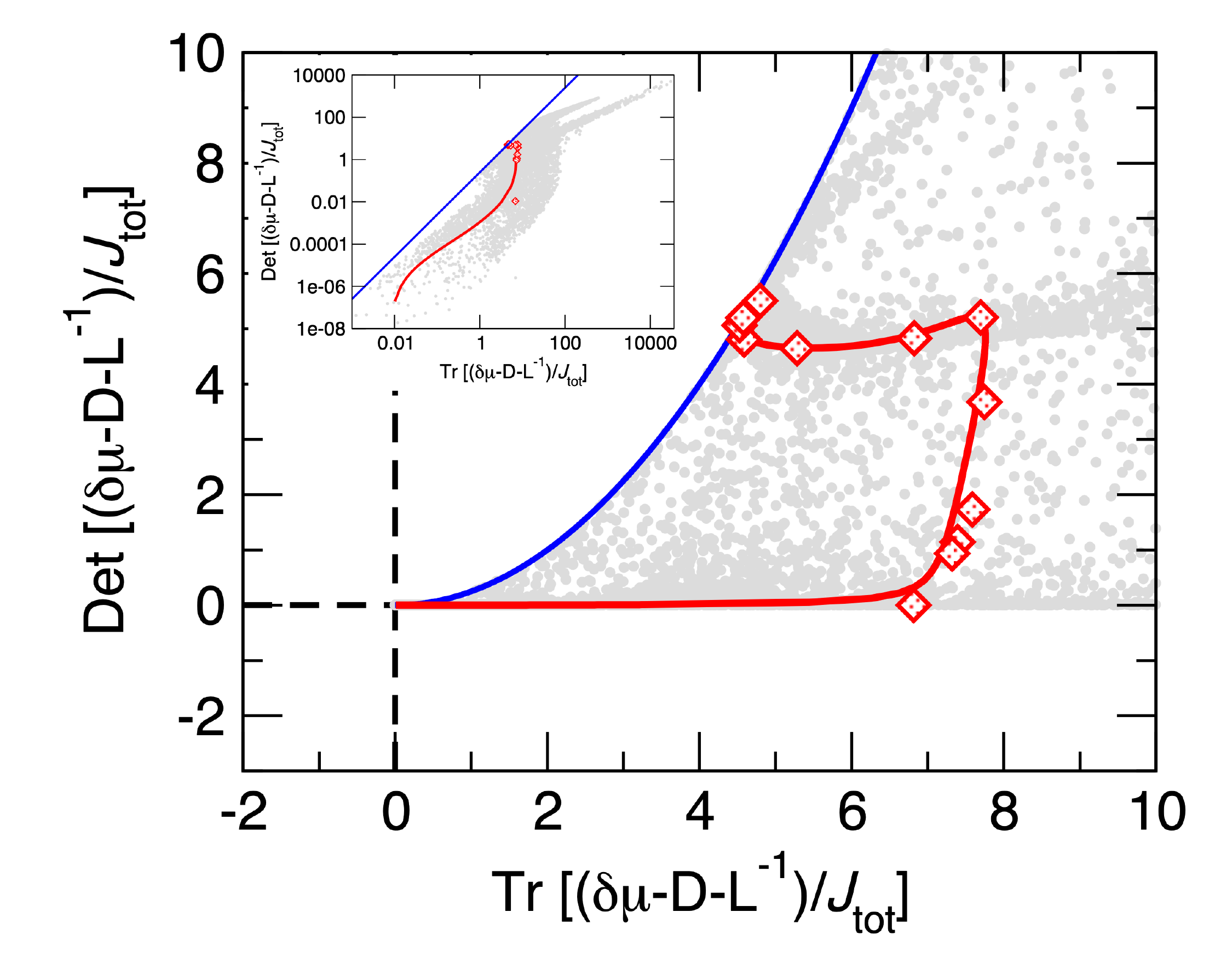}
\caption{Numerical verification of \cref{eq:central11}. The blue boundary marks the location of the inequality $\Tr[{\bf M}] \geq 2 \sqrt {\rm{Det}[{\bf M}]}$ for a two dimensional square matrix $\bf M$. The red diamonds are results from the non-equilibrium KMC simulations with the parameters used in Fig.\ \ref{fig:domain-lengths} and the red line is the theoretical mean field prediction for those same parameters. Gray dots are computed by constructing the matrices $\boldsymbol\delta\boldsymbol\mu$, $\bf D$, and $\bf L^{-1}$ using the the master equation results and computing the eigenvalues of  $({\boldsymbol\delta\boldsymbol\mu}-{\bf D}-{\bf L^{-1}})/J_{\rm tot}$ using Mathematica \cite{Mathematica} for randomly selected parameters from $L_{\beta, \rm eq} = [1,90000]$, $f_{\rm density,\beta}=[1,100] $ and $k_{\rm grow}=[0.001,100]$ nm/s, with all other parameters the same as the red line. The inset shows these two quantities for a wider range, with both axes in logarithmic scale.  We do not consider $k_{\rm grow} < 0.001$ nm/s due to limitations of numerical precision.
\cref{eq:central1}/\cref{eq:central11} provides strong constraints between the non-equilibrium forcing, morphology, and speed of growth. }
\label{fig:eigenvalues}
\end{figure}

We now define the matrices $\boldsymbol\delta\boldsymbol\mu$, $\bf D$ and $\bf L^{-1}$ precisely.
The matrix $\boldsymbol\delta\boldsymbol\mu$ depends on the non-equilibrium driving forces  $\delta\mu_{\alpha/\beta}\equiv \log dk_{\alpha/\beta}$ in \cref{eq:driving-in-rates-dk} and the average non-equilibrium domain lengths $L_{\alpha/\beta}$:
\begin{equation}
    \label{eq:centralMatdeltamu}
    \boldsymbol\delta\boldsymbol\mu=\left(\begin{matrix}\delta\mu_\alpha\gamma_1^\alpha+\delta\mu_{\beta}\gamma_2^\beta&\delta\mu_{\alpha}\gamma_3^\alpha+\delta\mu_{\beta}\gamma_3^\beta\\\delta\mu_{\alpha}\gamma_3^\alpha+\delta\mu_{\beta}\gamma_3^\beta&\delta\mu_{\alpha}\gamma_2^\alpha+\delta\mu_{\beta}\gamma_1^\beta\\\end{matrix}\right)
\end{equation}
where $\gamma_1^{\alpha/\beta}\equiv [(L_{\rm tot}-1)^2/(L_{\alpha/\beta}-1)+1]/L_{\rm tot}$, $\gamma_2^{\alpha/\beta}\equiv[1/(L_{\alpha/\beta}-1)+1]/L_{\rm tot}$, and $\gamma_3^{\alpha/\beta}\equiv -L_{\beta/\alpha}/[L_{\rm tot}(L_{\alpha/\beta}-1)]$. %Here $i$ stands for $\alpha/\beta$ while $j$ stands for $\beta/\alpha$. 
The matrix ${\bf D}$ depends on the non-equilibrium and equilibrium domain lengths of ABPs and is defined as 
\begin{equation}
    \label{eq:centralMatD}
    {\bf D}=\left(\begin{matrix}dp_{\alpha}\gamma_1^\alpha+dp_{\beta}\gamma_2^\beta+\epsilon& dp_{\alpha}\gamma_3^\alpha+dp_{\beta}\gamma_3^\beta+\epsilon\\dp_{\alpha}\gamma_3^\alpha+dp_{\beta}\gamma_3^\beta+\epsilon&dp_{\alpha}\gamma_2^\alpha+dp_{\beta}\gamma_1^\beta+\epsilon\\\end{matrix}\right)
\end{equation}
where $dp_{\alpha/\beta} \equiv (1/2) (\ln [(L_{\alpha/\beta}-1)/L_{\alpha/\beta}] -\ln [(L^{\rm eq}_{\alpha/\beta}-1)/L^{\rm eq}_{\alpha/\beta}])$ and $\epsilon \equiv (1/(2L_{\rm tot}))(\ln[(L^{\rm eq}_{\alpha}-1)/(L_{\alpha}-1)]+\ln[(L^{\rm eq}_{\beta}-1)/(L_{\beta}-1)])$. The $\bf{D}$ matrix only depends on the equilibrium and non-equilibrium morphologies of the bundle. 
$\bf{L^{-1}}$ is proportional to the inverse of the covariance matrix of fluxes and is computed as $\mathbf{L^{-1}} \equiv \lim_{t \rightarrow \infty} J_{\rm tot} {\mathbf M^{-1}}/t$, in which $\bf M$ has the elements $M_{ij}=\langle \delta J_i \delta J_j\rangle$ and $t$ is the time of growth. 
In \cref{fig:eigenvalues}, we numerically verify \cref{eq:central1}/\cref{eq:central11} for various parameter combinations. 
%(\frac{{\boldsymbol\delta\boldsymbol\mu}-{\bf D}}{J_{\rm tot}}) \cdot {\bf J} =

The equality in \cref{eq:central11} holds only when ${\boldsymbol\delta\boldsymbol\mu}-{\bf D}={\bf L^{-1}}$. In that case multiplying Eq.~\ref{eq:central1} by the column vector ${\bf J}$, containing the the average fluxes $J_{\alpha/\beta}$ computed using \cref{eq:mean-field-currents-domains} as $J_{\alpha/\beta}=J_{N-1,\alpha,N,\alpha/\beta}+J_{N-1,\beta,N,\alpha/\beta}$ (detailed in SI Section.~\ref{SI:FDT2}),
we readily obtain 
\begin{equation}
    \label{eq:central22}
      {\bf dk }-{\bf D[p]}=\tilde{{\bf L}}^{-1}\cdot {\bf J}
\end{equation}
where $\tilde{{\bf L}}^{-1}\equiv {\bf L}^{-1}/J_{\rm tot}$, $\mathbf{dk}$ is a column vector with elements $\log dk_{\alpha/\beta}$ (\cref{eq:driving-in-rates-dk}),  
and $\mathbf{D[p]}$ is a column vector with elements that are the relative entropies between the equilibrium and non-equilibrium domain morphologies of the two crosslinkers in the bundle (\cref{eq:dkdp,eq:dp}). 

\cref{eq:central22} can be viewed as an extension of the fluctuation dissipation relation to our non-equilibrium bundling and polymerization process. It relates the various driving forces ${\bf dk}$ and a relative entropic measure of the distance between the non-equilibrium and equilibrium structures, ${\bf D[p]}$, to the various observed fluxes ${\bf J}$ through the flux covariance matrix $\tilde{{\bf L}}^{-1}$.

The so called thermodynamic uncertainty relations (TUR)~\cite{barato2015thermodynamic,Gingrich2016,dechant2018multidimensional,horowitz2019thermodynamic,yan2019achievability} can also be readily derived from \cref{eq:central11}.  Specifically, \cref{eq:central11} implies that  $\tilde{{\bf J}}^T\cdot[{\boldsymbol\delta\boldsymbol\mu}-{\bf D}-{\bf L^{-1}}]\cdot\tilde{{\bf J}} \geq 0 $ for any vector $\tilde{{\bf J}}$. Hence, the so called multidimensional thermodynamic relation (MTUR)~\cite{dechant2018multidimensional} can be derived (SI Section~\ref{SI:MTUR}): 
\begin{equation}\label{eq:central2}
    J_{\rm tot} (\Delta\mu   - \varepsilon_{diss})=  2{\bf J^\intercal}\cdot( {\bf dk }-{\bf D[p]}) \geqslant  2{\bf J^\intercal}\cdot {\bf \tilde{{\bf L}}}^{-1} {\bf J}. 
\end{equation}

Our central result provides a connection between the microscopic  driving forces represented by $\boldsymbol\delta\boldsymbol\mu$ or $\bf{dk}$, the non-equilibrium structure of the bundle as encoded by matrix ${\bf D}$ or $\bf{D[p]}$, and the fluctuations of the various fluxes denoted by $\bf{L^{-1}}$ (obtained in the non-equilibrium steady state). Experimentally, it is possible to measure the fluxes of various bundling proteins, and the structure of the bundles. Then, one can use \cref{eq:central1} to bound the microscopic driving forces. These microscopic forces generally cannot be measured directly. Further, in non-equilibrium regimes where the $\bf{L^{-1}}$ matrix exhibits singular or close to singular behavior, our results suggest that the system might be insensitive to perturbations that tune the various microscopic driving forces, $\bf{dk}$. Our results suggest that the non-equilibrium bundling morphology can be effectively tuned away from such points. 

\begin{figure}
\includegraphics[width=\linewidth]{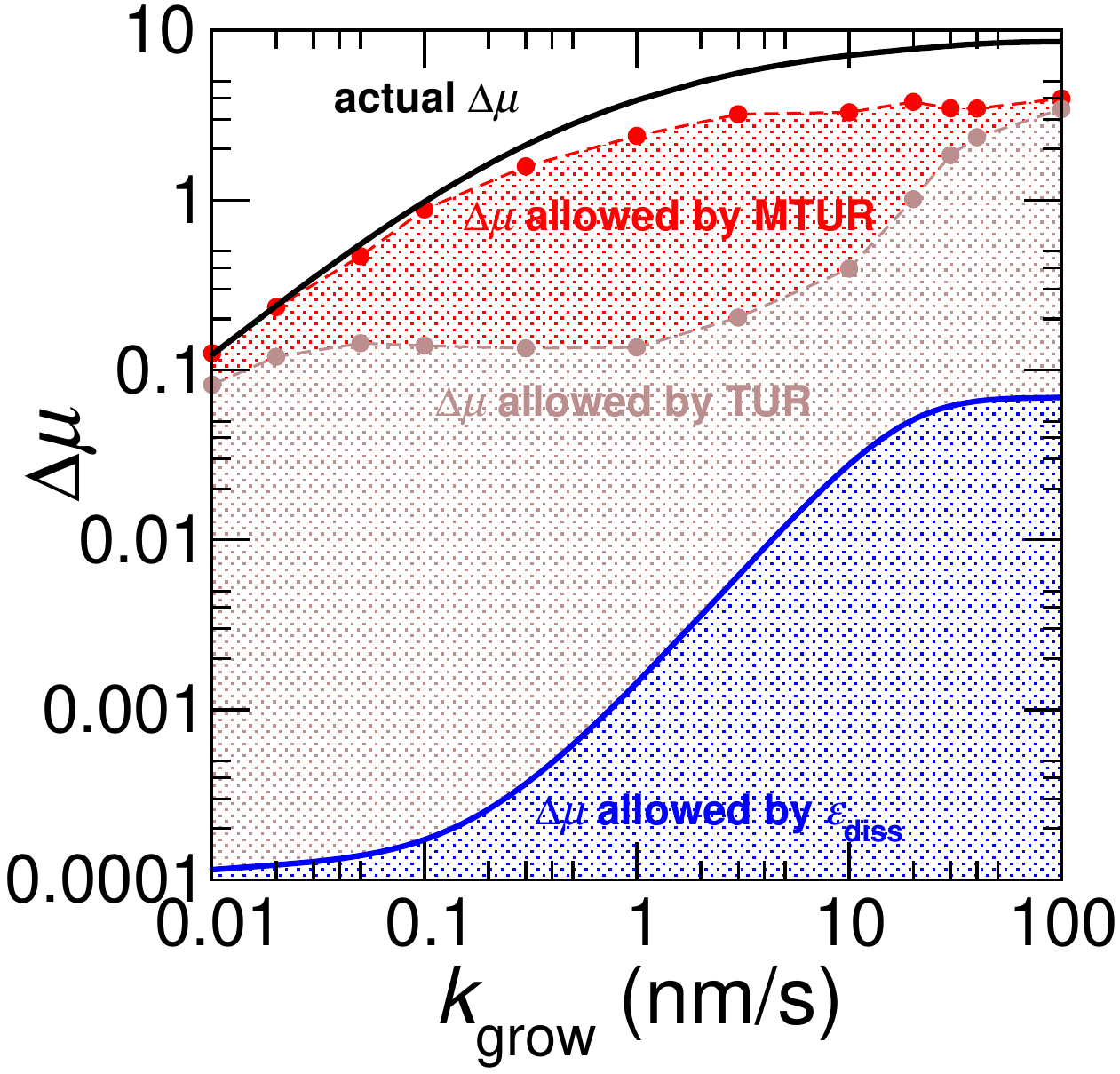}
\caption{Comparison of bounds on the non-equilibrium driving force $\Delta\mu$. The black line(\cref{eq:deltamu})is the actual driving force predicted by the master equation. The blue curve(\cref{eq:EP-secondlaw}) is the driving force required for morphology change. The brown (\cref{eq:TUR}) and red (\cref{eq:central2}) lines are the thermodynamic bounds computed from KMC simulations using TUR and MTUR. Each of the brown and red data points is generated with 500 independent KMC simulations, each of $10^7$ steps.  All parameters of the KMC simulations are the same as in \cref{fig:domain-lengths}.}
\label{fig:MTUR}
\end{figure}

Finally, \cref{eq:central1}/\cref{eq:central11} can also be used to assess the relative importance of accounting for the statistics of the individual fluxes. To do so, we use the thermodynamic uncertainty relations to derive a bound for the rate of entropy production in terms of the total flux, $J_{\rm tot}$: 
\begin{equation}\label{eq:TUR}
\Delta\mu  \geqslant   \varepsilon_{diss}  +\frac{2 \left\langle J_{\rm tot} \right\rangle}{t\left\langle\delta J_{\rm tot}^2\right\rangle}.
\end{equation}
Here, $t$ is the growth time of the bundle, $\langle J_{\rm tot}\rangle$ is the average total flux of adding ABPs to the bundle, and $\left\langle\delta J_{\rm tot}^2\right\rangle$ is its variance. In \cref{fig:MTUR}, we compare the performance of \cref{eq:TUR} (brown) with that of \cref{eq:EP-secondlaw} (blue). We see that the TUR bound is closer to the real driving $\Delta\mu$ compared with the second-law bound. Nevertheless, it still fails significantly at $k_{\rm grow} \approx 1$  nm/s, where it only recovers about 6\% of the actual driving. This implies that controlling the overall kinetics is not enough for facilitating the sorting of ABPs.

In \cref{fig:MTUR}, we also plot the MTUR bound (\cref{eq:central2}) using cumulants of fluctuations in the individual fluxes from KMC simulations. Although not perfect, this bound recovers at least 46\% of the actual driving $\Delta \mu$ for the full range of polymerization rates. The increasing gap between the MTUR bound and the actual driving is consistent with our central result in \cref{fig:eigenvalues} that Tr$[\boldsymbol\delta \boldsymbol\mu -\bf{D} - \bf{L^{-1}}]$ becomes further away from zero as microscopic driving becomes stronger and makes growth faster. 

The MTUR bound in \cref{eq:central2} is equivalent to considering a scalar observable $J_{\phi} = \cos\phi J_{\alpha}  + \sin\phi J_{\beta}$ and then maximizing $2 \left\langle J_{\phi}\right\rangle^2/(tJ_{\rm tot}\left\langle\delta J_{\phi}^2\right\rangle)$ by varying $\phi$ (SI~\cref{SI:MTUR}).
When $\tan\phi = 1$, we recover the TUR bound in \cref{eq:TUR}. Compared with this TUR bound, we find that the MTUR bound is mostly improved where the optimized $\tan\phi$ values deviate significantly from 1 (\cref{figs5}), and the fluxes are strongly correlated. Hence it is crucial to take into consideration the statistics of the individual fluxes. Indeed, the analytical expression of $\tan\phi$ (\cref{eqs19}) and the empirical results in \cref{figs6} and \cref{figs7} show that $\tan\phi$ is mainly governed by the ratio of the domain lengths, $L_\alpha/L_\beta$. In the regime $k_{\rm grow} \approx 0.1$ to $10$  nm/s, the domain lengths of the two ABPs differ significantly. We conclude that thermodynamic costs can be seriously masked if only the total flux instead of individual ones are resolved in regimes where the two ABPs display remarkably different sorting behavior. 

%We conclude that a large portion of the energy pumped into the system is invested to control the kinetics of this process, especially the strong correlation between fluxes that ultimately facilitates the sorting of ABPs. 
%Our work discovers that the correlation of fluxes cannot be neglected when reconstructing the non-equilibrium driving. %Using the same set of kinetic information, it is straightforward to quantify the resulting morphology of the network. 
%This would effectively connect the non-equilibrium driving to the morphology of biological networks.

%We validate this central result \cref{eq:central1} numerically and in KMC simulations. This bound is an extension of fluctuation dissipation relation that allows us to interpret how actin polymerization copes with the binding of ABPs to tune the non-equilibrium driving and to facilitate the sorting of ABPs. 
%In particular, we find that the non-equilibrium driving from actin polymerization is mostly consumed to regulate the correlation between fluxes of adding different types of ABPs to the bundle. This indicates that in order to estimate the microscopic driving force required for the bundling process far away from equilibrium, one need to measure carefully the individual fluxes of growth and account for the covariance between each pair of them.

\section{Conclusions}
In conclusion, we have derived a strong thermodynamic constraint relating the microscopic driving of a growing bundle (denoted by $\boldsymbol\delta\boldsymbol\mu$ in \cref{eq:central1}), the morphology of the bundle in its non-equilibrium steady state as described by the matrix $\bf{D}$, and the statistics of the rates of incorporation of crosslinkers as described by the matrix $\bf{L^{-1}}$. Our central results, which can be viewed as extensions of the fluctuation dissipation relations, also have practical applications. As an example, they potentially provide a route to estimate microscopic driving forces (contained in the ${\boldsymbol\delta\boldsymbol\mu}$ matrix) from experiments in which the various fluxes and morphologies are measured using microscopy and quantitative image analysis %or bulk spectroscopy~\cite{vavylonis2005actin}. 

While this current work is focused exclusively on the growth dynamics of bundled actin networks, we anticipate that the formalism presented here can be used in other contexts, such as the interplay between structure, speed, and non-equilibrium forcing in the growth dynamics of branched actin networks~\cite{liman2020role,bieling2016force,weichsel2010two}, the self-organization of other ABPs to distinct actin network architectures (e.g. networks initiated by formin or the Arp2/3 complex) \cite{kadzik2020f}, and the sorting of ABPs to distinct networks under confinement~\cite{bashirzadeh2020actin}.

\section{Acknowledgments}
This work was mainly supported by a DOE BES Grant DE-SC0019765 through funding to SV,YQ, MN.
YQ was also supported by a Yen Fellowship and the University of Chicago Materials Research Science and Engineering Center, which is funded by National Science Foundation under award number DMR-2011854. MN was also supported by a NSF Graduate research fellowship.   GMH was supported by National Institute of Health award R35 GM138312. ARD was supported by National Institute of Health award R35 GM136381. We thank Chatipat Lorpaiboon for help with the AFINES simulation software. Simulations were performed on the Midway cluster of the University of Chicago Research Computing Center.

\bibliographystyle{unsrt}
\bibliography{references}

\renewcommand\thesection{\arabic{section}}
\clearpage
\onecolumngrid

\section*{\LARGE Supporting Information}

\setcounter{figure}{0}
\setcounter{table}{0}
\setcounter{equation}{0}
\setcounter{page}{1}
\setcounter{section}{0}

\renewcommand{\theequation}{S\arabic{equation}} 
\renewcommand{\thepage}{S\arabic{page}} 
\renewcommand{\thesection}{S\arabic{section}}  
\renewcommand{\thetable}{S\arabic{table}}  
\renewcommand{\thefigure}{S\arabic{figure}}

\section{Kinetic Monte Carlo (KMC) Simulations}
\label{SI:KMC}
We simulate the growth of an actin bundle with two types of ABPs using KMC simulations. Two initial configurations are selected for each simulation: an actin bundle composed of 100 $\alpha$ type ABPs or a bundle composed of 100 $\beta$ type ABPs. We exclude this first 100 ABPs when measuring the final structure and find that domain lengths in the bundle are independent of its initial configuration in the parameter range we explore. For each step in the KMC simulations, we first identify which state the bundle tip is at, and all the possible forward and backward moves that can be initiated from this state. The KMC simulations are performed using the Gillespie algorithm \cite{gillespie1977exact}. 
We summarize the rates used in the KMC simulations in \cref{tables1}. To compute domain lengths in \cref{fig:domain-lengths}, we run one simulation of $S=10^6$ steps at each $k_{\rm grow}$ and measure domain lengths $L_{\alpha}$ and $L_{\beta}$ from the full bundle by counting the number of consecutive ABPs of the same type and averaging their lengths.
To generate the KMC data points in \cref{fig:eigenvalues} and \cref{fig:MTUR}, we run 500 simulations of $S=10^6$ steps at each actin polymerization rate $k_{\rm grow}>0.01$ nm/s; at $k_{\rm grow}=0.01$ nm/s, we run 5000 simulations of $S=2\times 10^8$ steps to ensure the convergence of the covariance of the fluxes. %In SI Section 9, we show that close to equilibrium ($k_{\rm grow}$ = $0.01$ nm/s), it requires more independent replications of longer KMC simulations to converge the $\bf{L}$ matrix.  %> 0.01$ nm/s. 
% table list the rates of KMC and master equation 
%We do not simulate at, or very close to, equilibrium because the bundles that result from $S=10^6$ steps are very short when $k_{\rm grow}<0.001$ nm/s, leading to large statistical uncertainties in the fluxes (\cref{fig:domain-lengths}).

%\section{2. The effect of actin polymerization $k_{\rm grow}$ encoded in the driving part of rates}

%We compute the driving part in rates for $\alpha$ and $\beta$ ABPs using \cref{eq:driving-in-rates-dk} with the same parameters of \cref{fig:domain-lengths}. When actin bundle stops growing at $k_{\rm grow} = 0$, the driving part $dk_{\alpha/\beta}$ is $1$ and the system is at equilibrium. The driving part increases to the maximum value $1+f_{\rm molecular} f_{\rm density}$ with increasing polymerization rate $k_{\rm grow}$. The maximum driving of $\beta$ ABP is higher than that of $\alpha$ ABP in fast growing bundles (\cref{figs1}). This is due to the higher value of \(f_{\rm molecular, \beta}\) compared with \(f_{\rm molecular, \alpha}\), which indicates that molecular details of ABPs determine their distinguished abilities in harnessing the non-equilibrium driving induced by fast bundle growth speed. ABPs in an equilibrium bundle can exchange with ABPs in reservoir for infinitely long time to minimize its free energy, so this molecular factor does not change the equilibrium bundle configurations. We conjecture that the coupling between molecular details of ABP  and actin polymerization is key in the sorting of ABPs in fast growing bundle if their densities are equal.

\section{Mean field master equation}
\label{SI:master}
\subsection{The backward rates in the master equation account for conditional probabilities during unbinding}

%The forward rates follow the form of \cref{eq:ratesk} and are the same for the KMC simulations and the master equation. We use the equilibrium part to account for the binding affinities of ABP and bending penalty of actins. In the next subsection, we derive the constraint on the equilibrium rates and equilibrium domain lengths. The driving part accounts for effect of actin polymerization $k_{\rm grow}$, ABP densities and their molecular structures (\cref{eq:driving-in-rates-dk}). 

\begin{figure*}
\centering
\includegraphics[scale=0.6, trim= 0cm 0cm 0cm 0cm, clip=true]{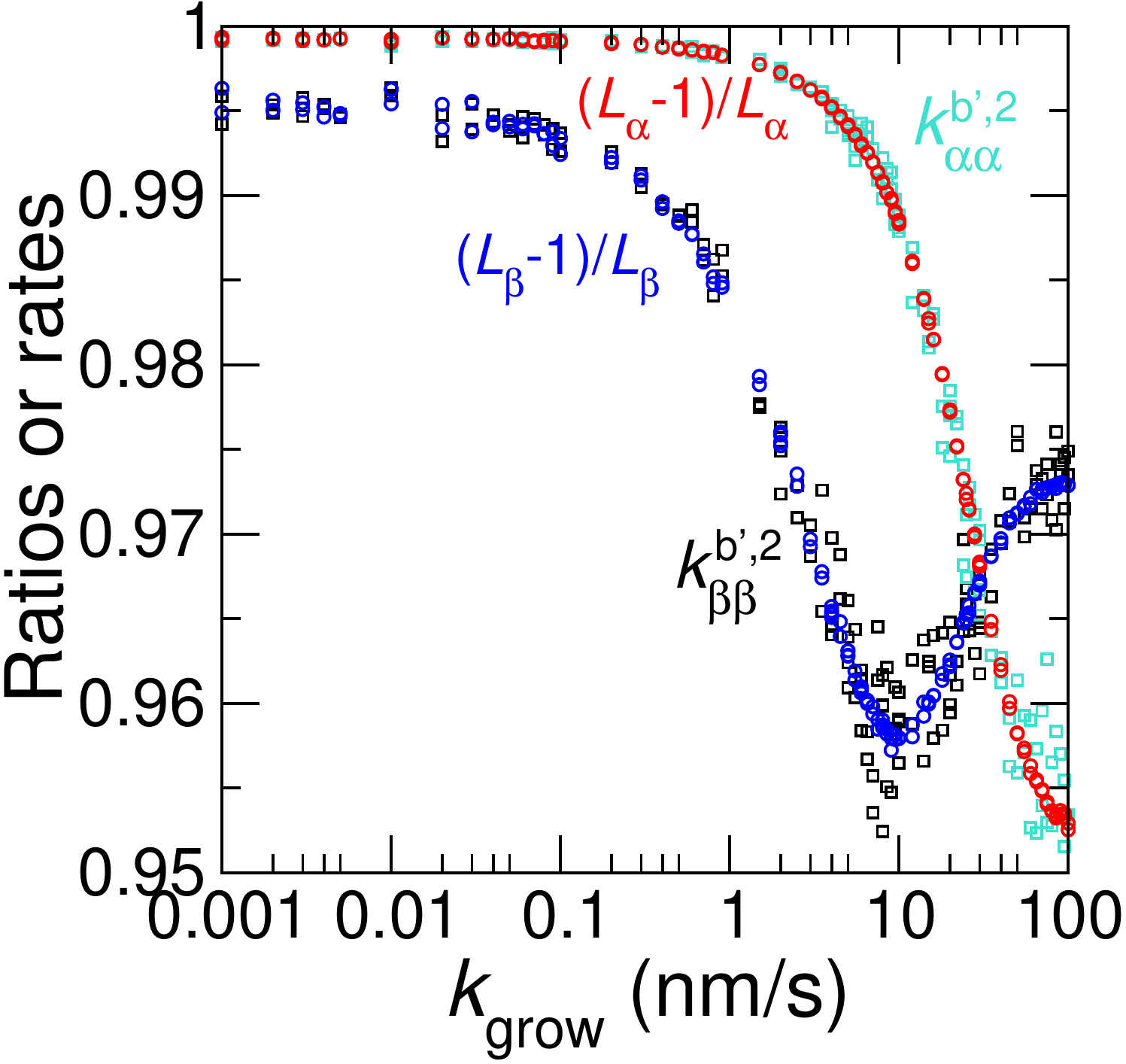}
\caption{ Validation of the backward rate expressions used in the master equation. Domain lengths ratio $(L_{\alpha}-1)/L_{\alpha}$ and $(L_{\beta}-1/L_{\beta}$  (blue and red circles) are computed from full bundle configurations in KMC simulations. The effective backward rates  $k^{b',2}_{\alpha\alpha}$ (cyan) and $k^{b',2}_{\beta\beta}$ (black) in KMC simulations are computed as $k^{b',2}_{\alpha\alpha} = N^{b,2}_{\alpha\alpha}/N^{b,2}_{\alpha\alpha}+N^{b,2}_{\beta\alpha}$ and $k^{b',2}_{\beta\beta} = N^{b,2}_{\beta\beta}/N^{b,2}_{\alpha\beta}+N^{b,2}_{\beta\beta}$. $N^{b,2}_{ij}$ is the number of occurrences of unbinding the second head of an ABP, where $i$ and $j$ are the types of the last two ABPs at the tip.}
\label{figs2}
\end{figure*}

We summarize the rates used in the master equation in \cref{tables1}.  
While in the KMC simulations the backward rates are unity,  in the master equation they include the conditional probabilities of finding the appropriate type of ABP during unbinding.  We thus express them in terms of the domain lengths in the bulk. To validate these expressions, we compute the effective backward rates in KMC simulations by counting the number of occurrences of each unbinding case. \cref{figs2} demonstrates that the expressions are quantitatively accurate.  

\begin{center}
\begin{table*}[htbp]
\centering

\caption{Forward and Backward rates for ABP in Master Equation(ME) and KMC Simulations}
\begin{tabular}{ >{\centering}p{2cm} |>{\centering}p{2cm}|>{\centering}p{2cm}|>{\centering}p{2cm}|>{\centering}p{2cm}|>{\centering}p{2cm}|>{\centering}p{1.5cm} }

\hline
 &Forward Rates & ME and KMC & Backward Rates(ME) & ME & Backward Rates(KMC) &KMC\tabularnewline
\hline
\multirow{4}{6em}{First head of ABP} & $k^{f,1}_{\alpha\alpha}$ & $ k^{f,1}_{\alpha\alpha}  =  k^{f,1}_{\beta\alpha}$ & $k^{b',1}_{\alpha\alpha}$ &  $1$ &$k^{b,1}_{\alpha\alpha}$ & $1$\tabularnewline

 & $k^{f,1}_{\alpha\beta}$    &   & $k^{b',1}_{\alpha\beta}$   & $1$ &$k^{b,1}_{\alpha\beta}$ & $1$ \tabularnewline
 
 &$k^{f,1}_{\beta\beta}$    & $ k^{f,1}_{\beta\beta}  = k^{f,1}_{\alpha\beta} $ & $k^{b',1}_{\beta\beta}$    &  $1$ & $k^{b,1}_{\beta\beta}$    & $1$\tabularnewline
 
 &$k^{f,1}_{\beta\alpha}$    &     & $k^{b',1}_{\beta\alpha}$   &  1&$k^{b,1}_{\beta\alpha}$ & $1$\tabularnewline

\hline
\multirow{4}{6em}{Second head of ABP} & $k^{f,2}_{\alpha\alpha}$    &   & $k^{b',2}_{\alpha\alpha}$   & $ (L_{\alpha}-1)/L_{\alpha}$   & $k^{b,2}_{\alpha\alpha}$ & $1$ \tabularnewline

&$k^{f,2}_{\beta\alpha}$    &    & $k^{b',2}_{\beta\alpha}$   & $1/L_{\alpha}$ &$k^{b,2}_{\beta\alpha}$   & $1$ \tabularnewline

&$k^{f,2}_{\beta\beta}$    &    & $k^{b',2}_{\beta\beta}$   & $ (L_{\beta}-1)/L_{\beta}$  &$k^{b,2}_{\beta\beta}$   & $1$ \tabularnewline
&$k^{f,2}_{\alpha\beta}$    &    & $k^{b',2}_{\alpha\beta}$   & $1/L_{\beta}$ & $k^{b,2}_{\alpha\beta}$ & 1 \tabularnewline
\hline

\end{tabular}
\label{tables1}
\end{table*}
\end{center}

%%%%

\subsection{Domain lengths and rates at equilibrium}
Given the rates in \cref{tables1}, we can solve the master equation (\cref{eq:master-eq}) at equilibrium to obtain constraints on the equilibrium rates. At equilibrium, all currents are zero, which results in the following relations between the equilibrium rates and equilibrium domain lengths, $L_{\alpha, \rm eq}$ and $L_{\beta, \rm eq}$.

\begin{equation}\label{eqs12}
\begin{split}
k^{f,1}_{\alpha\alpha,eq}k^{f,2}_{\alpha\alpha,eq} & = \frac{L_{\alpha, eq}-1}{L_{\alpha, eq}}\\
k^{f,1}_{\beta\beta,eq}k^{f,2}_{\beta\beta,eq}& = \frac{L_{\beta, eq}-1}{L_{\beta, eq}}\\
k^{f,1}_{\alpha\beta,eq}k^{f,2}_{\alpha\beta,eq}k^{f,1}_{\beta\alpha,eq}k^{f,2}_{\beta\alpha,eq} & = \frac{1}{L_{\alpha, eq}L_{\beta, eq}}\\
\end{split}
\end{equation}

Since the rate of binding of the first site does not depend on the filament separation, $k^{f,1}_{\alpha\alpha,eq} = k^{f,1}_{\beta\alpha,eq}$ and $k^{f,1}_{\beta\beta,eq} = k^{f,1}_{\alpha\beta,eq}$.  Consequently, there are six independent forward rates at equilibrium. We use the constraints above but split the third equation in \cref{eqs12} into the relations: 
\begin{equation}\label{eqs13}
\begin{split}
k^{f,1}_{\alpha\beta,eq}k^{f,2}_{\alpha\beta,eq} & = \frac{L_{\beta, \rm eq}-1}{L_{\beta, \rm eq}(L_{\alpha, \rm eq}-1)}\\
k^{f,1}_{\beta\alpha,eq}k^{f,2}_{\beta\alpha,eq} & = \frac{L_{\alpha, \rm eq}-1}{L_{\alpha, \rm eq}(L_{\beta, \rm eq}-1)}.
\end{split}
\end{equation}
An interpretation of \cref{eqs13} is that the effective bending penalty in switching from $\alpha$ to $\beta$ is $k^{f,2}_{\alpha\beta,eq}/k^{f,2}_{\beta\beta,eq} = 1/(L_{\alpha,eq}-1)$, and the bending penalty in switching from $\beta$ to $\alpha$ is $k^{f,2}_{\beta\alpha,eq}/k^{f,2}_{\alpha\alpha,eq} = 1/(L_{\beta,eq}-1)$. 
%This is consistent with the physical picture of actin bending penalty that the shorter the previous domain length is, the harder it is to bend the actin. 
We use these constraints on equilibrium rates to simplify the expression for the entropy production (see SI Section 4). Using \cref{eqs12} and \cref{eqs13}, we are able to represent the equilibrium condition with four parameters, $L_{\alpha, \rm eq}$, $L_{\beta, \rm \rm eq}$, $k^{f,1}_{\alpha\alpha,\rm eq}$, and $k^{f,1}_{\beta\beta,eq}$.

\subsection{Derivation of the self-consistency conditions}

Under non-equilibrium conditions, the currents are non-zero. We need an additional constraint on the domain lengths to solve the master equation for $L_{\alpha}$ and $L_{\beta}$. For this purpose, we derive the following self-consistency conditions, which are inspired by Ref.~\onlinecite{gaspard2014kinetics}. We consider a chain of ABPs with the sequence of \(\omega_{l}\omega_{l-1}...\omega_{0}\), where \(\omega_{i}\) represents the type of ABP at position $i$, and the crosslinker at the tip is index $i=0$. We write down an equation for the probability of finding this sequence at time \(t\), \(P^t(\omega_{l}\omega_{l-1}...\omega_{0})\):
\begin{equation} \label{eqs1}
\begin{split}
\frac{\partial P^t(\omega_{l}\omega_{l-1}...\omega_{0})}{\partial t} &= w^+_{\omega_0|\omega_1} P^t(\omega_{l}\omega_{l-1}...\omega_{1})\\
&+\sum_{Z} w^-_{Z|\omega_0} P^t(\omega_{l}|\omega_{l-1}\omega_{l-2}...\omega_{0})
P^t(\omega_{l-1}\omega_{l-2}...\omega_{0}|\omega_{l-2}\omega_{l-1}...\omega_{0}\omega_{Z}) P^t(\omega_{l-2}\omega_{l-1}...\omega_{0}\omega_{Z})\\  &-(w^-_{\omega_0|\omega_1}+\sum_{Z} w^+_{Z|\omega_0})  P^t(\omega_{l}|\omega_{l-1}\omega_{l-2}...\omega_{0})  P^t(\omega_{l-1}\omega_{l-2}...\omega_{0}), 
\end{split}
\end{equation}
where \(w^+_{\omega_i|\omega_j}\) is the rate of addition of \(\omega_{i}\) to \(\omega_{j}\) and \(w^-_{\omega_i|\omega_j}\) is the rate of removal of \(\omega_{i}\) to expose \(\omega_{j}\) at the tip. In what follows, we solve \cref{eqs1} at steady state:
\begin{equation} \label{eqs2}
\begin{split}
\frac{\partial P^t(\omega_{l}\omega_{l-1}\omega_{l-2}...\omega_{0})}{\partial t} = 0.
\end{split}
\end{equation}

We first substitute \cref{eqs1} into \cref{eqs2} and rearrange to express the possibility of having  \(\omega_{l}\) in addition to the sequence \(\omega_{l-1}\omega_{l-2}...\omega_{0}\).
\begin{equation} \label{eqs3}
\begin{split}
P^t(\omega_{l}|\omega_{l-1}\omega_{l-2}...\omega_{0})& = \frac{w^+_{\omega_0|\omega_1} P^t(\omega_{l}\omega_{l-1}...\omega_{1})}{(w^-_{\omega_0|\omega_1}+v_{\omega_0}) P^t(\omega_{l-1}\omega_{l-2}...\omega_{0})} \\
v_{\omega_0} &\equiv \frac{\sum_{Z} w^+_{Z|\omega_0}  P^t(\omega_{l-1}\omega_{l-2}...\omega_{0})-\sum_{Z} w^-_{Z|\omega_0}  P^t(\omega_{l-1}\omega_{l-2}...\omega_{0}\omega_{Z}) }{P^t(\omega_{l-1}\omega_{l-2}...\omega_{0})}
\end{split}
\end{equation}
Physically, \(v_{\omega_0}\) is the growing rate at the tip with crosslinker type \(\omega_0\). By further rearranging \cref{eqs3} and summing over $\omega_0$, we obtain the following expression. 
\begin{equation}
\label{eqs4}
\sum_{\omega_0}w^+_{\omega_0|\omega_1} P^t(\omega_{l}\omega_{l-1}...\omega_{1})- \sum_{\omega_0}w^-_{\omega_0|\omega_1}P^t(\omega_{l}\omega_{l-1}...\omega_{0})
  = \sum_{\omega_0} v_{\omega_0}  P^t(\omega_{l}|\omega_{l-1}\omega_{l-2}...\omega_{0}) P^t(\omega_{l-1}\omega_{l-2}...\omega_{0})
\end{equation}
The left side of \cref{eqs4} is \(v_{\omega_1}  P^t(\omega_{l}\omega_{l-1}...\omega_{1})\), so \cref{eqs4} becomes 
\begin{equation} \label{eqs5}
\begin{split}
v_{\omega_1}P^t(\omega_{l}\omega_{l-1}...\omega_{1}) =  \sum_{\omega_0}& v_{\omega_0}  P^t(\omega_{l-1}\omega_{l-2}...\omega_{0})P^t(\omega_{l}|\omega_{l-1}\omega_{l-2}...\omega_{0}) 
\end{split}
\end{equation}
\cref{eqs5} is equivalent to Eq. 34 in ref.\  \cite{gaspard2014kinetics}.
We use \(\Gamma_{0}\) to represent the configuration \(\omega_{l-1}\omega_{l-2}...\omega_{0}\) and \(\Gamma_{1}\) to represent the configuration \(\omega_{l}\omega_{l-2}...\omega_{1}\). Imagining that we focus on the configuration with a fixed length \(l\) from the tip of the chain \(\omega_{0}\) and  moving backwards one position each time, \cref{eqs5} provides a connection between the probability of finding \(\Gamma_{1}\)  and the probability of finding \(\Gamma_{0}\). We can write \cref{eqs5} in matrix form as
%\begin{equation}
%        W(\Gamma_{1}|\Gamma_{0})= 
%\begin{cases}
%    P^t(\omega_{l}|\omega_{l-1}\omega_{l-2}...\omega_{0}),& \text{if } \Gamma_{1}(m)=\Gamma_{0}(m-1)   \forall m >1 \\
%   0,              & \text{otherwise}
%\end{cases}
%\end{equation}
%where $\Gamma_i(m)$ is the $m^{\rm th}$ element of the sequence $\Gamma_i$. 
\begin{equation} \label{eqs6}
\begin{split}
P^t(\Gamma_{1}) = W(\Gamma_{1}|\Gamma_{0})P^t(\Gamma_{0})
\end{split}
\end{equation}
where $W(\Gamma_{1}|\Gamma_{0}) = P^t(\omega_{l}|\omega_{l-1}\omega_{l-2}...\omega_{0})$ for consistent sequences $\Gamma_0$ and $\Gamma_1$.
Consequently, applying $W$ $N$ times, we have
\begin{equation} \label{eqs6}
\begin{split}
P^t(\Gamma_{N}) = W^NP^t(\Gamma_{0}).
\end{split}
\end{equation}

At steady state, the eigenvector of \(W\) with eigenvalue 1 is the probability $\bar{P}(\Gamma)$ of finding a particular sequence $\Gamma$ in the bulk. From \cref{eqs5} we can now immediately see that
\begin{equation} \label{eqs7}
\begin{split}
\bar{P}(\Gamma_\infty) =\frac{v_{tip}(\Gamma_\infty)}{\bar{v}}P^t(\Gamma_\infty),
\end{split}
\end{equation}
where $\Gamma_\infty$ denotes a sequence in the bulk,  ${\bar{v}}$ is the total growth rate at the tip for all configurations, and $v_{tip}(\Gamma_i) = v_{\omega_i}$.  
In what follows, we use \cref{eqs7} to derive the self-consistency conditions specific to our system. We choose two neighboring ABPs \(\omega_{m}\omega_{m-1}\) for the configuration size, where $m$ ranges from 1 to \(l\). There are four configurations: $\omega_{m,\alpha},\omega_{m-1,\alpha}$, 
\(\omega_{m,\beta},\omega_{m-1,\alpha}\), \(\omega_{m,\alpha},\omega_{m-1,\beta}\), and \(\omega_{m,\beta},\omega_{m-1,\beta}\). We refer to the term on the right hand side of \cref{eqs7} as currents in the main text. 

\begin{equation}\label{eqs8}
\begin{split}
v_{tip}(\omega_{m,\alpha},\omega_{m-1,\alpha}){P}^t(\omega_{m,\alpha},\omega_{m-1,\alpha}) \equiv & J_{\omega_{m,\alpha},\omega_{m-1,\alpha}}
\\
= &k^{f,1}_{\alpha\alpha}P(\alpha) - k^{b,1}_{\alpha\alpha}P(\alpha\alpha*)
= k^{f,2}_{\alpha\alpha}P(\alpha\alpha*) - k^{b,2}_{\alpha\alpha}P(\alpha)\\
v_{tip}(\omega_{m,\beta},\omega_{m-1,\beta}){P}^t(\omega_{m,\beta},\omega_{m-1,\beta}) \equiv & J_{\omega_{m,\beta},\omega_{m-1,\beta}} \\
= &k^{f,1}_{\beta\beta}P(\beta) - k^{b,1}_{\beta\beta}P(\beta\beta*)
= k^{f,2}_{\beta\beta}P(\beta\beta*) - k^{b,2}_{\beta\beta}P(\beta)\\
v_{tip}(\omega_{m,\alpha},\omega_{m-1,\beta}){P}^t(\omega_{m,\alpha},\omega_{m-1,\beta}) \equiv& J_{\omega_{m,\alpha},\omega_{m-1,\beta}} \\
= &k^{f,1}_{\alpha\beta}P(\alpha) - k^{b,1}_{\alpha\beta}P(\alpha\beta*)
= k^{f,2}_{\alpha\beta}P(\alpha\beta*) - k^{b,2}_{\alpha\beta}P(\beta)\\
v_{tip}(\omega_{m,\beta},\omega_{m-1,\alpha}){P}^t(\omega_{m,\beta},\omega_{m-1,\alpha}) \equiv& J_{\omega_{m,\beta},\omega_{m-1,\alpha}} \\
= &k^{f,1}_{\beta\alpha}P(\beta) - k^{b,1}_{\beta\alpha}P(\beta\alpha*)
= k^{f,2}_{\beta\alpha}P(\beta\alpha*) - k^{b,2}_{\beta\alpha}P(\alpha)\\
\end{split}
\end{equation}
where $J_{\omega_{m,i},\omega_{m-1,j}}$ is the current of ABPs of type $j$ binding after ABPs of type $i$. $J_{\rm tot}$ is the sum of the four currents:
\begin{equation}\label{eqs9}
\begin{split}
\bar{v} &\equiv J_{\rm tot} =J_{\omega_{m-1,\alpha},\omega_{m,\alpha}}+J_{\omega_{m-1,\alpha},\omega_{m,\beta}}+J_{\omega_{m-1,\beta},\omega_{m,\beta}}+J_{\omega_{m-1,\beta},\omega_{m,\alpha}}.
\end{split}
\end{equation}

By combining \cref{eqs7} and \cref{eqs8}, we can obtain the following self-consistency condition, 
\begin{equation}\label{eqs10}
\begin{split}
J_{\omega_{m-1,\alpha},\omega_{m,\alpha}}&=\bar{P}^t(\omega_{m,\alpha},\omega_{m-1,\alpha}) J_{\rm tot}\\
J_{\omega_{m-1,\beta},\omega_{m,\beta}}&= \bar{P}^t(\omega_{m,\beta},\omega_{m-1,\beta})J_{\rm tot}\\
J_{\omega_{m-1,\alpha},\omega_{m,\beta}}=J_{\omega_{m-1,\beta},\omega_{m,\alpha}} &= \bar{P}^t(\omega_{m,\alpha},\omega_{m-1,\beta})J_{\rm tot}.
\end{split}
\end{equation}

 The current $J_{\omega_{m-1,\alpha},\omega_{m,\beta}}$ is identical to the current  $J_{\omega_{m-1,\beta},\omega_{m,\alpha}}$ because for every switch of a domain of $\alpha$ crosslinkers there is a switch to a domain of $\beta$ crosslinkers.  $\bar{P}^t(\omega_{m},\omega_{m-1})$ is the probability of finding a specific configuration \(\omega_{m}\omega_{m-1}\) in the bulk; it can be expressed in terms of the domain lengths as follows. 
\begin{equation}\label{eqs11}
\begin{split}
&\bar{P}^t(\omega_{m,\alpha},\omega_{m-1,\alpha}) = \frac{L_{\alpha}-1}{L_{\alpha}+L_{\beta}} \\
&\bar{P}^t(\omega_{m,\beta},\omega_{m-1,\beta}) = \frac{L_{\beta}-1}{L+L_{\beta}} \\
&\bar{P}^t(\omega_{m,\alpha},\omega_{m-1,\beta}) = \bar{P}^t(\omega_{m,\beta},\omega_{m-1,\alpha}) = \frac{1}{L_{\alpha}+L_{\beta}} 
\end{split}
\end{equation}

\begin{figure*}
\centering
\includegraphics[scale=0.5, trim= 0cm 0cm 0cm 0cm, clip=true]{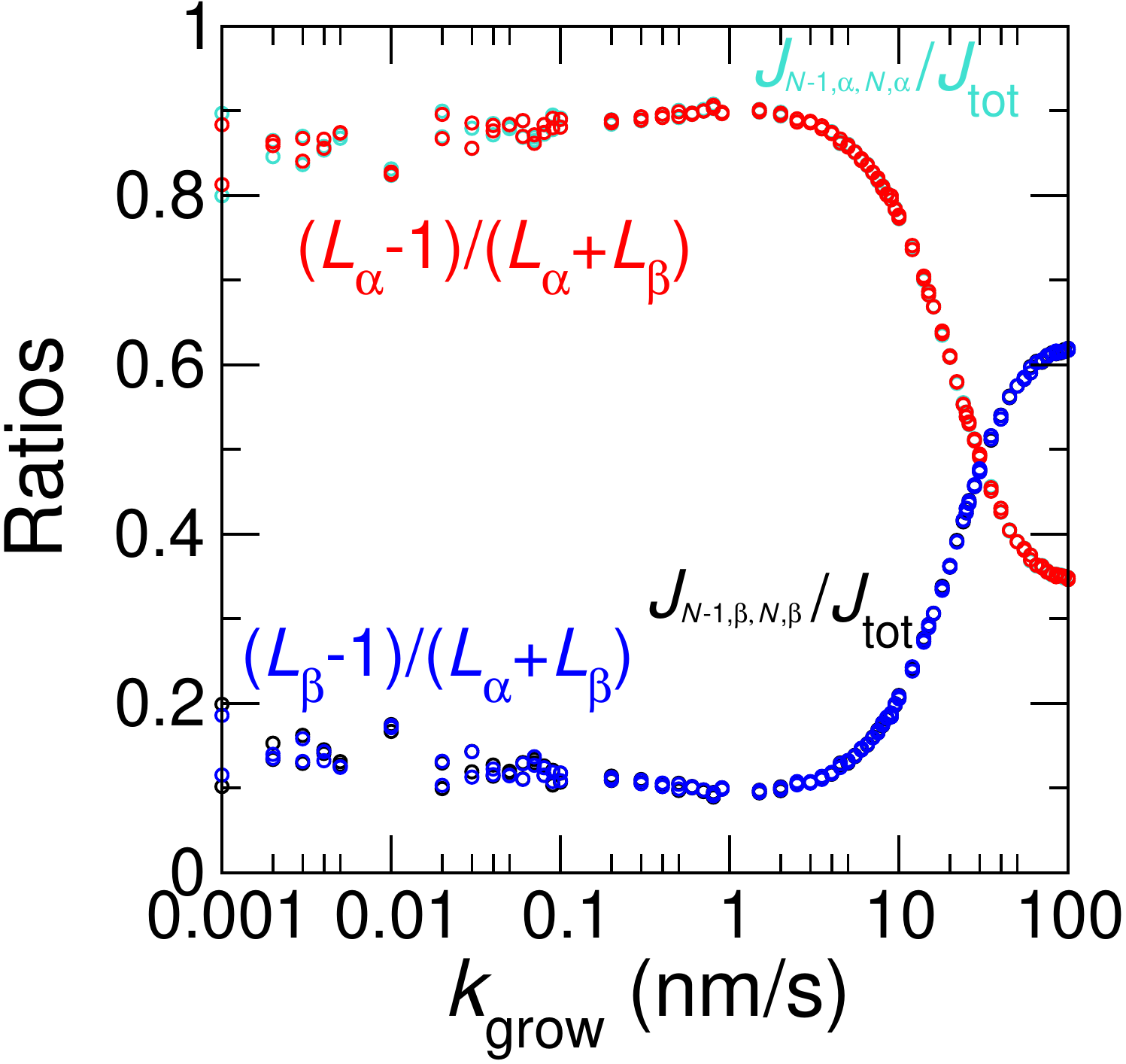}
\caption{Comparison of currents and domain length ratios from KMC simulations at various actin polymerization rates. Both currents and domain lengths are measured from KMC simulations of $10^6$ steps. The simulation parameters are the same as those used in generating \cref{fig:domain-lengths}.}
\label{figs3}
\end{figure*}

Using \cref{eqs10} and \cref{eqs11}, we obtain \cref{eq:mean-field-currents-domains}. Note that in main text we omit the $\omega$'s in the subscripts of the currents for clarity. The self-consistency conditions are validated by KMC simulations (\cref{figs3}). 

\subsection{Domain lengths predicted by master equation}
\label{SI:Domain}
Since the equilibrium parameters and the driving parts of the rates are determined, we now use the self-consistency condition above to solve the master equation (\cref{eq:master-eq}) for the domain lengths $L_{\alpha}$ and $L_{\beta}$ under non-equilibrium conditions. 
\begin{equation}\label{eqs14}
\begin{split}
L_{\beta} = \frac{(L_{\alpha}-1)k^{f,2}_{\beta\alpha}-k^{f,2}_{\alpha\alpha}(L_{\alpha}( k^{f,2}_{\beta\alpha}+1) k^{f,1}_{\alpha\alpha}+L_{\alpha}-1)}{k^{f,2}_{\beta\alpha} k^{f,1}_{\alpha\alpha} (Lk^{f,2}_{\alpha\alpha}((1+k^{f,2}_{\alpha\beta})k^{f,1}_{\alpha\alpha}-(L_{\alpha}-1)k^{f,2}_{\alpha\beta} k^{f,1}_{\beta\beta})-(L_{\alpha}-1)(1+k^{f,2}_{\alpha\beta}+Lk^{f,2}_{\alpha\beta}k^{f,1}_{\beta\beta}))}\\
L_{\alpha} = \frac{(L_{\beta}-1)k^{f,2}_{\alpha\beta}-k^{f,2}_{\beta\beta}(L_{\beta} (k^{f,2}_{\alpha\beta}+1) k^{f,1}_{\beta\beta}+L_{\beta}-1)}{k^{f,2}_{\alpha\beta} k^{f,1}_{\beta\beta} (Sk^{f,2}_{\beta\beta}((1+k^{f,2}_{\beta\alpha})k^{f,1}_{\beta\beta}-(L_{\beta}-1)k^{f,2}_{\beta\alpha} k^{f,1}_{\alpha\alpha})-(L_{\beta}-1)(1+k^{f,2}_{\beta\alpha}+Sk^{f,2}_{\beta\alpha}k^{f,1}_{\alpha\alpha}))} 
\end{split}
\end{equation}
%Although we can obtain the domain lengths $L_{\alpha}$ and $L_{\beta}$ analytically from \cref{eqs14} in terms of rates, the resulting expressions are complicated. Instead, 
We compute the domain lengths from \cref{eqs14} numerically and compare them with those measured from KMC simulations in \cref{fig:domain-lengths}. We solve \cref{eqs14} using standard numerical routines in MATHEMATICA. The quality of the numerical solutions depends on the effectiveness of our mean field approach. In general, we find that the mean field approximations behind \cref{eqs14} works well far from equilibrium but perform poorly close to equilibrium.

\section{Derivation of Eq.\ (1)}
\label{SI:FDT}
\begin{figure}
\centering
\includegraphics[scale=0.5]{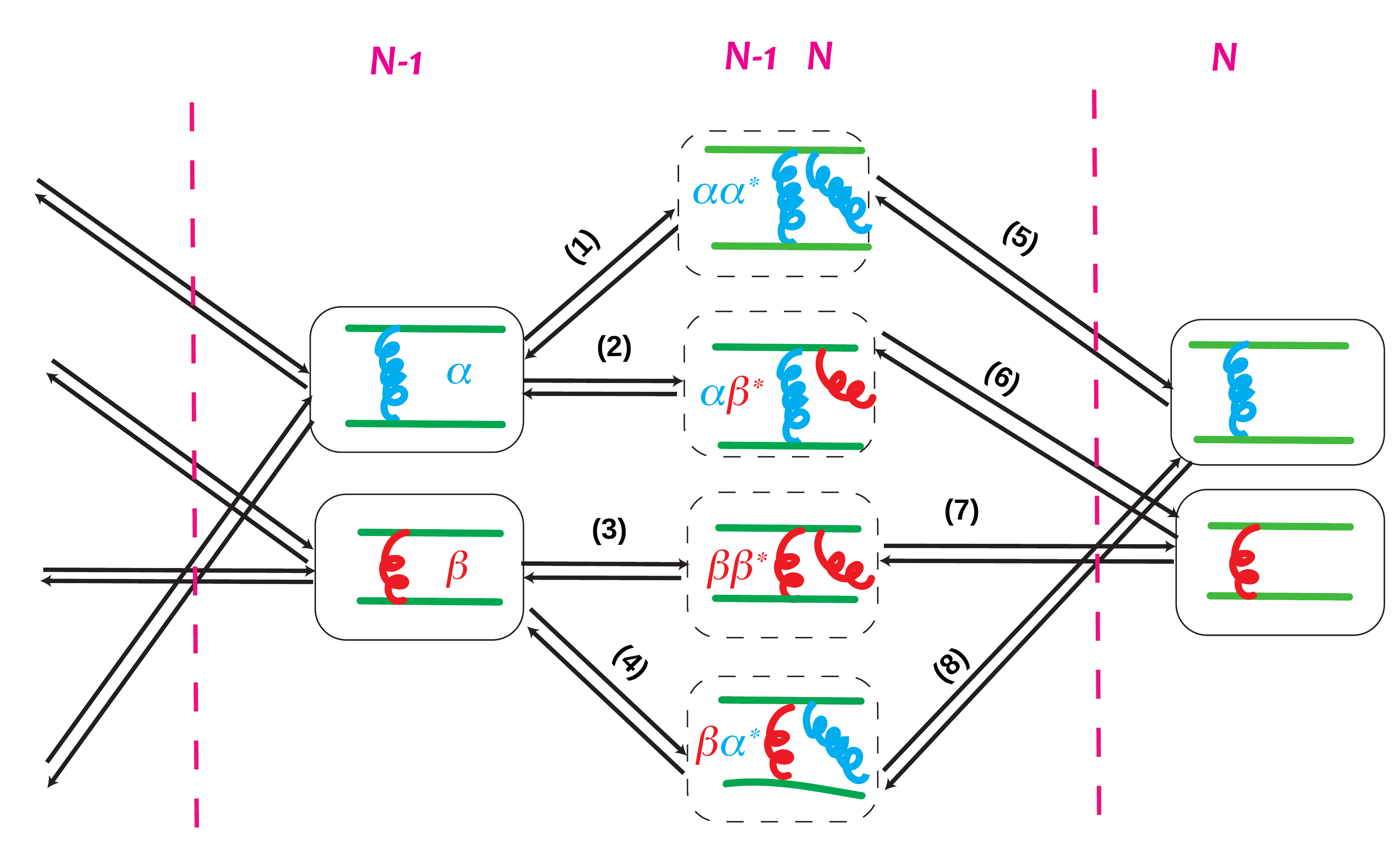}
\caption{Schematic showing all the possible transitions of the actin bundle system. We index the graph edges from 1 to 8.}
\label{figs8}
\end{figure}

In this section, we derive \cref{eq:central1} of the main text. Our derivation closely follows those in Refs.~\onlinecite{Gingrich2016,yan2019achievability}. As in the main text, we focus of experimentally accessible currents $J_\alpha$ and $J_\beta$ related to the net rate at which the different crosslinkers are assimilated into the bundle. In particular, using the notation in Fig.~\ref{figs8}, we have  
\begin{equation}\label{eqs27}
\begin{split}
    J_\alpha&=[j(1)+j(4)+j(5)+j(8)]/2\\
    J_\beta&=[j(2)+j(3)+j(6)+j(7)]/2,
\end{split}
\end{equation}
where we use $j(\epsilon)$ to denote the current along the edge $\epsilon$. 

\cref{eqs27} can be written in matrix form as \begin{equation}\label{eqs20}
    J(n)=\sum_\epsilon j(\epsilon) { d}(\epsilon,n)
\end{equation}
 where ${ d}(n,\epsilon)$ are elements in a matrix $\bf d$ that describes how each of the generalized currents depends on the edge currents. For the specific case described above in \cref{eqs27}, the
 $\mathbf{d}$ matrix can be written as:
\begin{equation}
\label{dmatrix}
        \mathbf{d}^T=\frac{1}{2}
    \begin{pmatrix}
    1&0&0&1&1&0&0&1\\
    0&1&1&0&0&1&1&0
    \end{pmatrix}
\end{equation}

Before proceeding further, we outline the structure of the proof. We attempt to derive \cref{eq:central1} by deriving constraints on the fluctuations of $J_\alpha$ and $J_\beta$ (or equivalently $J(n)$ in \cref{eqs20}). To do so we first express $J(n)$ in terms of the edge currents $\tilde{j}(\epsilon)$ such that $\tilde {j}(\epsilon)$ are guaranteed to satisfy current conservation, and they satisfy $J(n)=\sum_\epsilon \tilde j(\epsilon) d(\epsilon,n)$. Then, we use the findings of Ref.~\onlinecite{Gingrich2016} which showed that the large deviation rate functions associated with the fluctuations in the various edge currents satisfy the following inequality, 
\begin{equation}\label{eqs22}
    I(\textbf{j})\leq\sum_\epsilon \left (j(\epsilon)-j^{\pi}(\epsilon)\right)^2\frac{\sigma^\pi(\epsilon)}{4[j(\epsilon)]^2},
\end{equation}
where $I(\textbf{j})$ is the rate function of the edge currents and $\sigma^\pi(\epsilon)$ is the entropy production of edge $\epsilon$. 
Finally, by writing  \cref{eqs22} in terms of the generalized currents $J(n)$, we obtain our central result.

We now proceed by first ``inverting" \cref{eqs20} and defining a set of edge currents:
\begin{equation}\label{eqs21}
    \tilde{j}(\epsilon)=\sum_n J(n) G(n,\epsilon)
\end{equation}
where ${\bf G}$ is a pseudoinverse of the matrix ${\bf d}$. Note that since  ${\bf G}$ and ${\bf d}$ are pseudoinverses of one another, we have 
\begin{equation}\label{eqs21xx}
    \sum_\epsilon \tilde{j}(\epsilon)  d(\epsilon,k)=\sum_{n,\epsilon} J(n) G(n,\epsilon)d(\epsilon,k)=\sum_{n,\epsilon,\epsilon^\prime} j(\epsilon^\prime) { d}(\epsilon^\prime,n) G(n,\epsilon)d(\epsilon,k)=\sum_{\epsilon^\prime} j(\epsilon^\prime)d(\epsilon^\prime,k)=J(k)
\end{equation}

If the set of currents $\tilde{{\bf j}}$ satisfy current conservation \textendash we will construct a matrix {\bf G} below in Eq.~\ref{eqs29} that satisfies this constraint \textendash we can use the arguments in Ref.~\onlinecite{Gingrich2016,yan2019achievability} to substitute \cref{eqs21} into \cref{eqs22} and obtain a bound on the large deviation rate function associated with the generalized currents, $I({\bf J})$
\begin{equation}\label{eqs23}
    I(\textbf{J})\leq\sum_\epsilon \left (\sum_n J(n) G(n,\epsilon)-J^{\pi}(\epsilon)G(n,\epsilon)\right)^2\frac{\sigma^\pi(\epsilon)}{4[j(\epsilon)]^2}.
\end{equation}
Now let's consider the system at steady state, with $\textbf{J}^\pi$ denoting the vector of average generalized currents and $\bf{L}$ denoting the covariance of generalized currents. The rate function $I(\textbf{J})$ can be expanded around the average generalized currents, $\textbf{J}^\pi$, as 
\begin{equation}\label{expand}
    I(\textbf{J})\approx I(\textbf{J}^\pi)+D\left[I(\textbf{J})|\textbf{J}^\pi\right]\cdot\mathbf{\tilde{J}}+\frac{1}{2}\mathbf{\tilde{J}}^T\cdot H\left[I(\textbf{J})|\textbf{J}^\pi\right]\cdot\mathbf{\tilde{J}}
\end{equation}
Here $\mathbf{\tilde{J}}$ is a vector with elements $(J(n)-J^{\pi}(n))$, and $D\left[I(\textbf{J})|\textbf{J}^\pi\right]$ is the vector containing the derivatives of the rate function $I(\textbf{J})$ with respect to the current $\textbf{J}$. Since the rate function is at its minimum at $\textbf{J}^\pi$, $I(\textbf{J}^\pi)$ and $D\left[I(\textbf{J})|\textbf{J}^\pi\right]$ equal to 0.  $H\left[I(\textbf{J})|\textbf{J}^\pi\right]$ is the Hessian matrix of $I(\textbf{J})$ evaluated at $\textbf{J}^\pi$ which can be related to the covariance matrix $\bf{L}$ by  \cite{ellis1985}
\begin{equation}\label{eqs24}
    H\left[I(\textbf{J})|\textbf{J}^\pi\right]=\bf{L^{-1}}.
\end{equation}
\cref{expand} and \cref{eqs24} then allow us to rewrite \cref{eqs23} as
\begin{equation}\label{eqs25}
    \mathbf{\tilde{J}}^T\cdot \mathbf{L}^{-1}\cdot\mathbf{\tilde{J}}\leq \mathbf{\tilde{J}}^T\cdot \mathbf{GSG}\cdot\mathbf{\tilde{J}}
\end{equation}
Here $\mathbf{G}$ is the matrix form of $G(n,\epsilon)$ with $n$ indicating the row and $\epsilon$ indicating the column,  $\mathbf{G^T}$ is its transpose and $\mathbf{S}$ is a diagonal matrix with elements $\sigma(\epsilon)/2|j(\epsilon)|^2$. 
Since \cref{eqs23} is valid for any arbitrary fluctuation about the mean,  $\mathbf{GSG}^T-\mathbf{L}^{-1}$ has to be positive semi-definite, which is to say that all of its eigenvalues have to be non-negative. For a $2\cross2$ matrix, this is equivalent to
\begin{equation}\label{eqs26}
    \rm{Tr}(\mathbf{GSG}^T-\mathbf{L}^{-1})\geq 0,\quad \rm{Det}(\mathbf{GSG}^T-\mathbf{L}^{-1})\geq 0.
\end{equation}

%explain here this only works when fluxes are conserved, and how we choose matrix G 
To finish the proof, we now need to fix a matrix $G$. The choice of $G$ is guided by the fact that the inequality in \cref{eqs23} is only valid when the currents $\tilde{\bf j}$ are conserved  currents, which means the sum of the currents that go through each node is zero. This gives the following constraints on the edge currents $\tilde{j}_{i}$ (with the notation as specified in Fig.~\ref{figs8}). 
\begin{equation}\label{eqs28}
    \tilde{j}(1)=\tilde{j}(5),\quad  \tilde{j}(3)=\tilde{j}(7),\quad \tilde{j}(2)=\tilde{j}(4)=\tilde{j}(6)=\tilde{j}(8)
\end{equation}
These constraints require that the elements in ${\bf G}$ have the following relations.
\begin{equation}\label{eq:constraints}
    G(i,1)=G(i,5),\quad G(i,3)=G(i,7), \quad G(i,2)=G(i,4)=G(i,6)=G(i,8)
\end{equation}
where $i = 1$ or $2$ are the first and second row of ${\bf G}$ and represent the contributions of $\alpha$ and $\beta$ components of the generalized currents to each edge currents.

The following  ${\bf G}$ matrix written in terms of $L_{\alpha}$ and $L_{\beta}$ satisfies all the constraints and is also a psuedoinverse of the ${\bf d}$ matrix:
\begin{equation}\label{eqs29}
    {\bf G}=\frac{1}{(L_\alpha+L_\beta)}
    \begin{pmatrix}
    L_\alpha+L_\beta-1&1&-1&1&L_\alpha+L_\beta-1&1&-1&1\\
    -1&1&L_\alpha+L_\beta-1&1&-1&1&L_\alpha+L_\beta-1&1
    \end{pmatrix}
\end{equation}

Substituting the above equation into $\mathbf{GSG}^T$ and collecting the terms with $\ln{dk_{\alpha/\beta}}$ gives us the matrix $\boldsymbol\delta\boldsymbol\mu$ in the main text. The rest is the matrix $\mathbf{D}$ in main text.

\section{Derivation of the linear response like identity}
\label{SI:FDT2}
A fluctuation dissipation like form can also be obtained as in \cref{eq:central22} of the main text. In this section, we derive \cref{eq:central22} using the matrix $\mathbf{GSG}^T$ we determined in \cref{SI:FDT}. First, we multiply the matrix $\mathbf{GSG}^T$ by the vector $\mathbf{J}$ on both sides:
\begin{equation}
    \label{entropyprodfromG}
    \mathbf{J}^T\mathbf{GSG}^T\mathbf{J}=\mathbf{j}^T\mathbf{S}\mathbf{j}=\dot{\sigma}/2
\end{equation}
With $\mathbf{J}^T=(J_\alpha,J_\beta)$ and $\mathbf{j}^T=(j_1,j_2,...,j_8)$
The entropy production can be further rewritten as: 
\begin{equation}\label{eqs33}
\dot{\sigma}=J_{\rm tot} (\Delta\mu - \varepsilon_{diss}) = 2\mathbf{J}^T\left(\mathbf{dk}-\mathbf{D[p]}\right)
\end{equation}
where the microscopic force vector $\textbf{dk}$ and the relative entropy term ${\bf D[p]}$ are defined as
\begin{equation}\label{eq:dkdp}
\begin{split}
\mathbf{dk} &= \begin{pmatrix} \log{dk_{\alpha}}\\
\log{dk_{\beta}}
\end{pmatrix}\\
{\bf D[p]} &= \begin{pmatrix}
D(P(\alpha)||P^{eq}(\alpha))\\D(P(\beta)||P^{eq}(\beta)) \end{pmatrix}
\end{split}
\end{equation}
with
\begin{equation}\label{eq:dp}
\begin{split}
 D(P(i)||P^{eq}(i))= &\frac{L_{i}-1}{2L_{i}} (\log\frac{L_{i}-1}{L_{i}}- \log\frac{L_{i, \rm eq}-1}{L_{i, \rm eq}})
+\frac{1}{2L_{i}} (\log\frac{1}{L_{i}}- \log\frac{1}{L_{i, \rm eq}})
\end{split}
\end{equation}

We then arrive at:

\begin{equation}
    \label{fluctuationdissform}
    \mathbf{J}^T\mathbf{GSG}^T\mathbf{J}=\mathbf{J}^T\left(\boldsymbol\delta\boldsymbol\mu-\mathbf{D}\right)\mathbf{J}=\mathbf{J}^T\left(\mathbf{dk}-\mathbf{D[p]}\right)
\end{equation}
So when $\boldsymbol\delta\boldsymbol\mu-\mathbf{D} = \mathbf{L}^{-1}$, we obtain \cref{eq:central22} in the main text.

%\end{document}
%To validate \cref{eq:central1} numerically, we construct the matrices ${\boldsymbol\delta\boldsymbol\mu}$, ${\bf D}$, and $\bf L^{-1}$ using the expressions in the main text (\cref{eq:centralMatdeltamu,eq:centralMatD}). All variables are computed directly from input parameters and analytical expressions in \cref{eq:ratesk,eqs14}.  Then we compute the eigenvalues of  $({\boldsymbol\delta\boldsymbol\mu}-{\bf D}-{\bf L^{-1}})/J_{\rm tot}$ using the following procedures with Mathematica \cite{Mathematica}. 
% need to explain more of this part
%the covariance of fluxes from master equation by taking the second  derivative of the eigenvalue of matrix $W$ in \cref{eq:master-eq}.

\section{The thermodynamic bound for driving is improved by considering individual currents instead of the total current in the TUR}
\label{SI:MTUR}
We use \cref{eqs9} to re-write \cref{eq:EP-secondlaw}:
\begin{equation}\label{eqs15}
\begin{split}
\dot\sigma = J_{\rm tot}\left(\frac{L_{\alpha}-1}{L_{\alpha}+L_{\beta}}\log\frac{k^{f,1}_{\alpha\alpha}k^{f,2}_{\alpha\alpha}}{k^{b,1}_{\alpha\alpha}k^{b,2}_{\alpha\alpha}}+
\frac{L_{\beta}-1}{L_{\alpha}+L_{\beta}}\log\frac{k^{f,1}_{\beta\beta}k^{f,2}_{\beta\beta}}{k^{b,1}_{\beta\beta}k^{b,2}_{\beta\beta}}
+\frac{1}{L_{\alpha}+L_{\beta}}\log\frac{k^{f,1}_{\alpha\beta}k^{f,2}_{\alpha\beta}k^{f,1}_{\beta\alpha}k^{f,2}_{\beta\alpha}}{k^{b,1}_{\beta\beta}k^{b,2}_{\beta\beta}k^{b,1}_{\beta\alpha}k^{b,2}_{\beta\alpha}}\right)
\end{split}
\end{equation}

We combine \cref{eqs12,eqs13,eqs15,eqs16} to obtain the following expression for entropy production in terms of $\Delta\mu$ and $\varepsilon_{diss}$ in \cref{eq:deltamu} and \cref{eq:ediss}
\begin{equation}\label{eqs16}
\begin{split}
\dot\sigma = J_{\rm tot}(\Delta\mu -\varepsilon_{diss})
\end{split}
\end{equation}

We first plot the second law bound and the TUR bound (\cref{eq:TUR}) in \cref{fig:MTUR}. The TUR bound is much better than the second law bound because it encodes the kinetic information of the process. It deviates from the real driving at intermediate polymerization rate $k_{\rm grow}= 1$ nm/s. For further improvement, we can adapt the MTUR bound~\cite{dechant2018multidimensional} to this process. We define the fluxes of adding $\alpha$ and $\beta$ ABPs as $J_{\alpha} = J_{\omega_{m-1,\alpha},\omega_{m,\alpha}} + J_{\omega_{m-1,\beta},\omega_{m,\alpha}}$  and  $J_{\beta} = J_{\omega_{m-1,\alpha},\omega_{m,\beta}} + J_{\omega_{m-1,\beta},\omega_{m,\beta}}$.The MTUR bound given by \cref{eq:central2} is the same as considering the scalar observable $J_{tot'} = \cos\phi J_{\alpha}  + \sin\phi J_{\beta}$ and then maximizing $2 \left\langle J_{tot'}\right\rangle^2/tJ_{\rm tot}\left\langle\delta J_{tot'}^2\right\rangle$ by varying $\phi$. We demonstrate below the derivation and plot the $\tan \phi$ values that optimize the bound as a function of $k_{\rm grow}$.

Inserting the expression of $J_{tot'}$ into the bound of  $2 \left\langle J_{tot'}\right\rangle^2/tJ_{\rm tot}\left\langle\delta J_{tot'}^2\right\rangle$, we obtain
\begin{equation}\label{eqs17}
\begin{split}
\Delta\mu  \geqslant   \varepsilon_{diss} +\frac{2 (J_{\alpha}+\tan\phi J_{\beta})^2}{t J_{\rm tot} (\langle\delta {J_{\alpha}}^2\rangle +\tan^{2}\phi\langle\delta {J_{\beta}}^2\rangle + 2 \tan\phi \langle\delta {J_{\alpha}} \delta {J_{\beta}}\rangle) }
\end{split}
\end{equation}

We take the derivative of \cref{eqs17} against $\tan\phi$ and find that the maximum value is achieved when 
\begin{equation}\label{eqs18}
\begin{split}
\tan\phi = \frac{-\langle\delta {J_{\alpha}} \delta {J_{\beta}}\rangle J_{\alpha} + \langle\delta {J_{\alpha}}^2\rangle J_{\beta}}{-\langle\delta {J_{\alpha}} \delta {J_{\beta}}\rangle J_{\beta} + \langle\delta {J_{\beta}}^2\rangle J_{\alpha}}.
\end{split}
\end{equation}

We substitute \cref{eqs18} for $\tan\phi$ in \cref{eqs17} and obtain the following MTUR bound for the bundling process. 
\begin{equation}\label{MTUR-full}
\begin{split}
\Delta\mu  \geqslant   \varepsilon_{diss}  +\frac{2 (J_{\alpha}^2 \langle\delta {J_{\beta}}^2\rangle - 2\langle\delta {J_{\alpha}} \delta {J_{\beta}}\rangle J_{\alpha} J_{\beta} +J_{\beta}^2 \langle\delta {J_{\alpha}}^2\rangle)}{t J_{\rm tot} ( -\langle\delta {J_{\alpha}} \delta {J_{\beta}}\rangle^2 + \langle\delta {J_{\alpha}}^2\rangle\langle\delta {J_{\beta}}^2\rangle) }.
\end{split}
\end{equation}

\begin{figure}
\centering
\includegraphics[scale=0.5, trim= 0cm 0cm 0cm 0cm, clip=true]{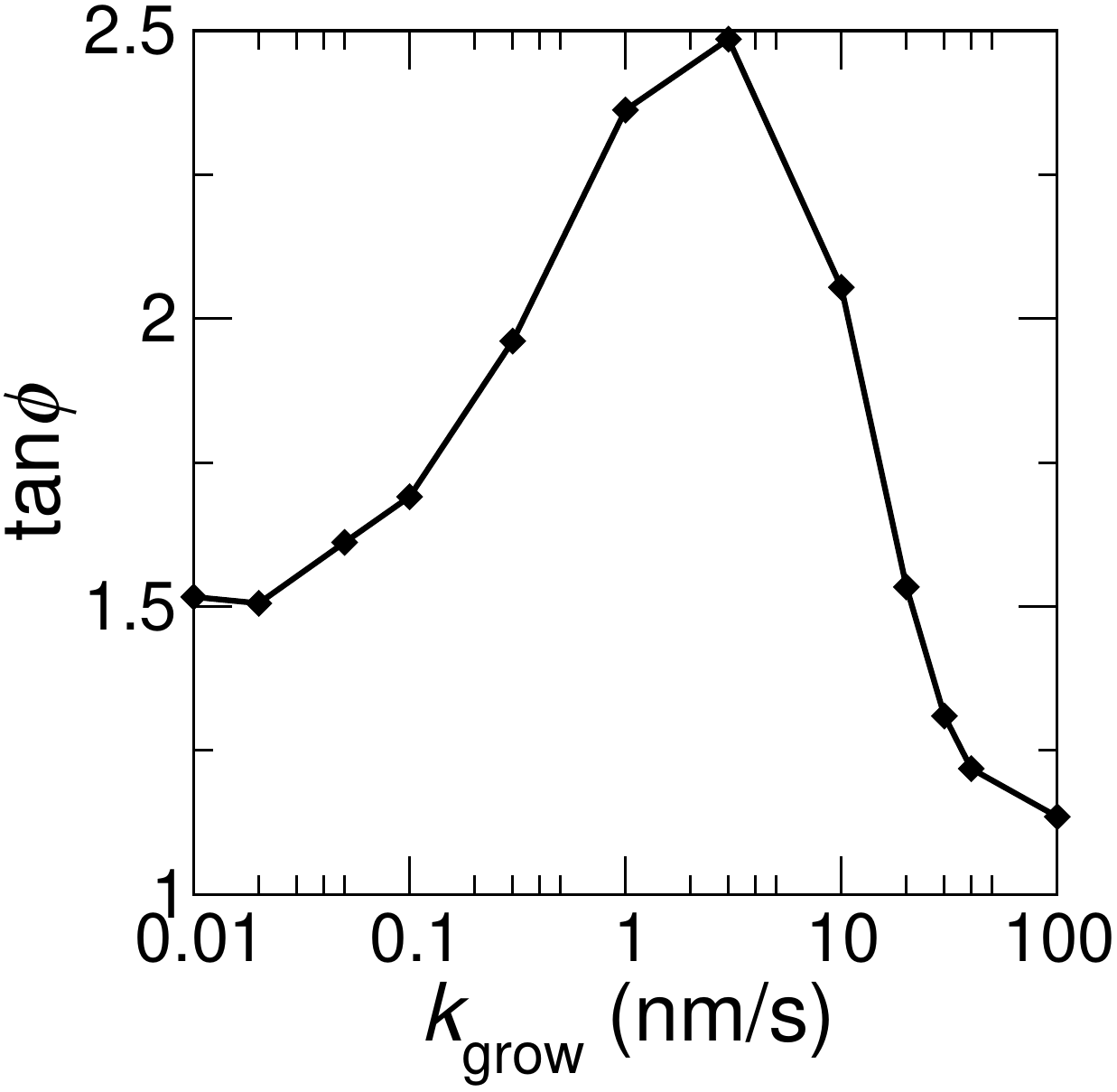}
\caption{Coefficient $\tan\phi$ at various actin polymerization rates $k_{\rm grow}$. The parameters are the same as in \cref{fig:MTUR}.}
\label{figs5}
\end{figure}

\begin{figure}
\centering
\includegraphics[scale=0.5, trim= 0cm 0cm 0cm 0cm, clip=true]{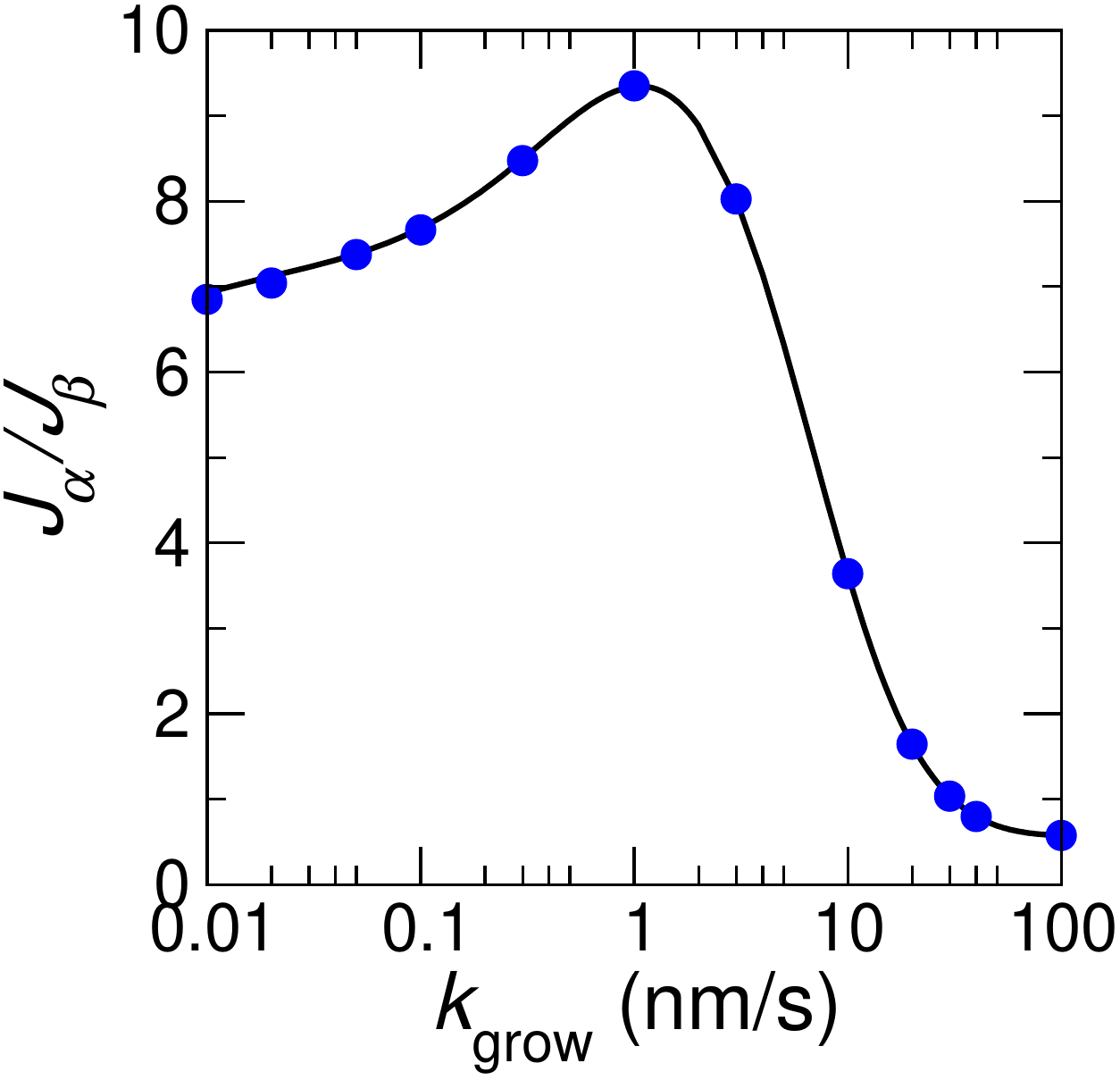}
\caption{The ratio between the two currents $J_{\alpha}/J_{\beta}$ computed from KMC simulations (blue points) and predicted by the master equation (black curve). }
\label{figs6}
\end{figure}

This is equivalent to \cref{eq:central2}, where we simplify the expression using
\begin{equation}
\bf{J} = \begin{pmatrix} J_{\alpha} \\ J_{\beta} \end{pmatrix}, 
\bf{dk} = \begin{pmatrix} \log dk_{\alpha} \\ \log dk_{\beta} \end{pmatrix}\ \textrm{and}\ \bf{D[p]} = \begin{pmatrix} D(P(\alpha)||P^{eq}(\alpha)) \\ D(P(\beta)||P^{eq}(\beta)) \end{pmatrix}.
\end{equation}
\cref{figs5} shows the value of $\tan\phi$ in \cref{eqs18} at various $k_{\rm grow}$. This coefficient is obtained by maximizing the bound in \cref{eqs17} at each data point.  It provides information about the strength of the correlation between the two currents. We compare this bound with the one given by the TUR (\cref{eq:TUR}) in \cref{fig:MTUR} and discuss their performance in the main text.
%shows that the performance of the MTUR is better than the TUR, especially in the polymerization rate range $k_{\rm grow}$ where $\tan\phi$ deviates from 1. This indicates that it is necessary to include the correlation between individual currents in computing the thermodynamic bound. 

\begin{figure}
\centering
\includegraphics[scale=0.5, trim= 0cm 0cm 0cm 0cm, clip=true]{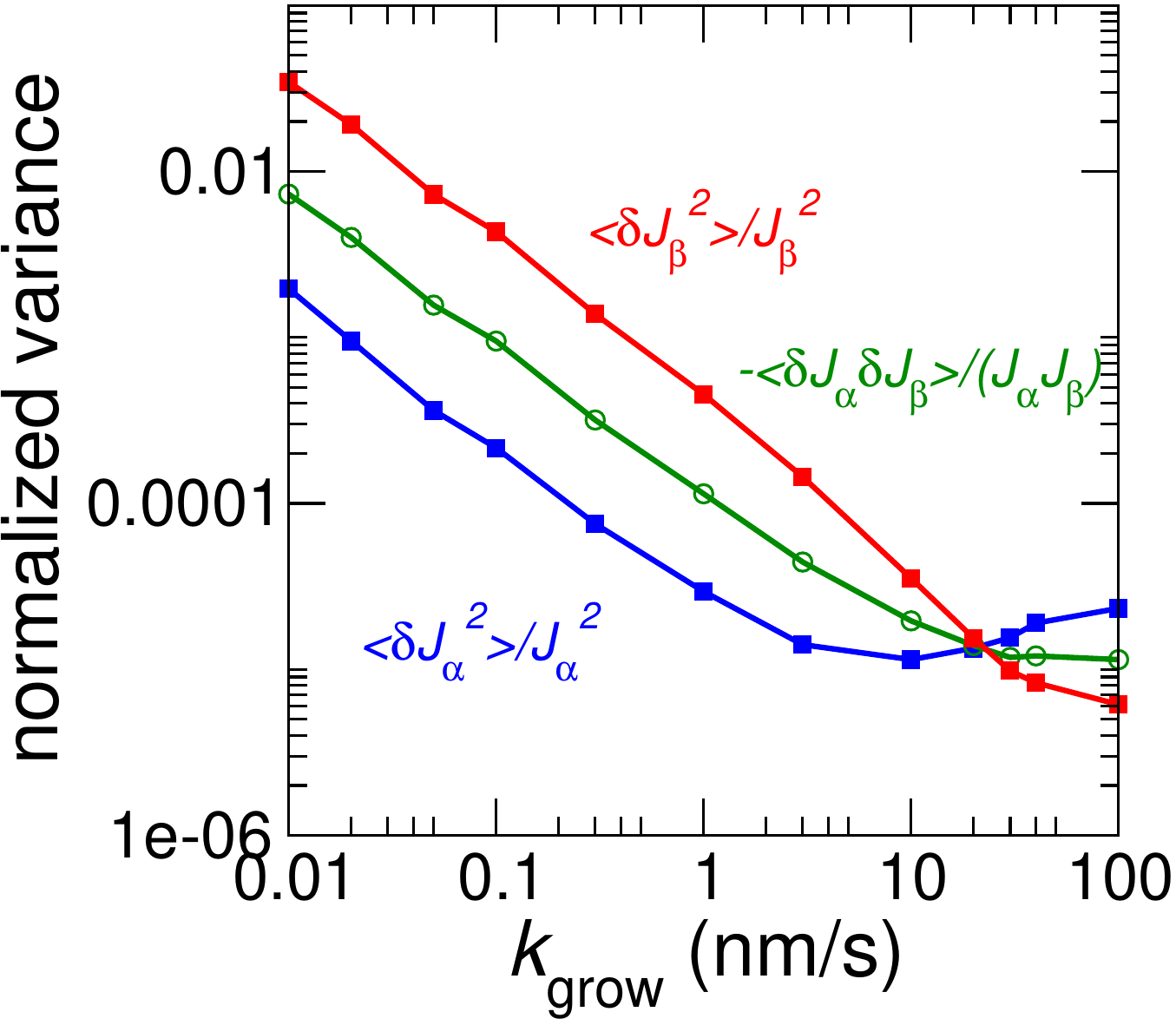}
\caption{Normalized variance and covariance computed from KMC simulations. Blue and red points are the variance normalized by the corresponding current ($\langle\delta {J_{\beta}}^2\rangle/J_{\beta}^2$ and $\langle\delta {J_{\alpha}}^2\rangle/J_{\alpha}^2$) computed from simulations. Green points are the covariance normalized by the magnitude of the two currents ($-\langle\delta {J_{\alpha}} \delta {J_{\beta}}\rangle/J_{\alpha}J_{\beta}$).}
\label{figs7}
\end{figure}

To further analyze how the variance and covariance of fluxes modulate the value of $\tan\phi$ and the amount of driving used to maintain the correlation between fluxes, we rearrange the expression of $\tan\phi$ as follows. 
\begin{equation}\label{eqs19}
\begin{split}
\tan\phi = \frac{J_{\alpha}}{J_{\beta}}  \left[ \frac{\frac{-\langle\delta {J_{\alpha}} \delta {J_{\beta}}\rangle}{J_{\alpha}J_{\beta}} +\frac{\langle\delta {J_{\alpha}}^2\rangle}{{J_{\alpha}}^2}}{\frac{-\langle\delta {J_{\alpha}} \delta {J_{\beta}}\rangle}{J_{\alpha}J_{\beta}} +\frac{\langle\delta {J_{\beta}}^2\rangle}{{J_{\beta}}^2}}\right]
\end{split}
\end{equation}
\cref{figs5} shows that  the ratio between currents  $J_{\alpha}/J_{\beta}$ has the same non-monotonic dependence on actin polymerization rate as $\tan\phi$. This is consistent with \cref{eqs19}. On the other hand, the normalized variance and covariance of currents $-\langle\delta {J_{\alpha}} \delta {J_{\beta}}\rangle/J_{\alpha}J_{\beta}$, $\langle\delta {J_{\alpha}}^2\rangle/J_{\alpha}^2$ and $\langle\delta {J_{\beta}}^2\rangle/J_{\beta}^2$ have similar orders of magnitude for a given $k_{\rm grow}$ (\cref{figs7}), suggesting that the value of $\tan\phi$ is weakly dependent on the second term on the right side of \cref{eqs19}. We conclude that $\tan\phi$, as well as the strength of the correlation between fluxes is mainly determined by the ratio between the two currents.

\end{document}